\documentclass[fleqn,usenatbib]{mnras}

\usepackage{newtxtext,newtxmath}

\usepackage[T1]{fontenc}

\DeclareRobustCommand{\VAN}[3]{#2}
\let\VANthebibliography\thebibliography
\def\thebibliography{\DeclareRobustCommand{\VAN}[3]{##3}\VANthebibliography}

\usepackage{graphicx}	
\usepackage{amsmath}	
\usepackage{xcolor}
\usepackage{float}
\usepackage{siunitx}
\usepackage{setspace}
\usepackage{geometry}
\usepackage{graphicx}
\usepackage{natbib}
\usepackage{pdfpages}
\usepackage{subcaption}
\usepackage[toc,page]{appendix}
\usepackage{threeparttable}

\newcommand{\Halpha}{H$\rm \alpha$ }
\newcommand{\Porb}{$P_{\rm orb}$}

\title{Growth and Dissipation of Be Star Discs in Misaligned Binary Systems}

\author[M. Suffak et al.]{
M. Suffak,$^{1}$\thanks{E-mail: msuffak@uwo.ca}
C.E. Jones,$^{1}$
A.C. Carciofi$^{2}$
\\
$^{1}$Department of Physics and Astronomy, Western University, London, ON N6A 3K7, Canada\\
$^{2}$Instituto de Astronomia, Geof\'isica e Ci\'encias Atmosf\'ericas, Universidade de S\~ao Paulo, Brazil\\
}

\date{Accepted XXX. Received YYY; in original form ZZZ}

\pubyear{2021}

\begin{document}
\label{firstpage}
\pagerange{\pageref{firstpage}--\pageref{lastpage}}
\maketitle

\begin{abstract}
We use a three-dimensional smoothed particle hydrodynamics code to simulate growth and dissipation of Be star discs in systems where the binary orbit is misaligned with respect to the spin axis of the primary star. We investigate six different scenarios of varying orbital period and misalignment angle, feeding the disc at a constant rate for 100 orbital periods, and then letting the disc dissipate for 100 orbital periods. During the disc growth phase, we find that the binary companion tilts the disc away from its initial plane at the equator of the primary star before settling to a constant orientation after 40 to 50 orbital periods. While the mass-injection into the disc is ongoing, the tilting of the disc can cause material to reaccrete onto the primary star prematurely. Once disc dissipation begins, usually the disc precesses about the binary companion's orbital axis with precession periods ranging from 20 to 50 orbital periods. In special cases we detect phenomena of disc tearing, as well as Kozai-Lidov oscillations of the disc. These oscillations reach a maximum eccentricity of about 0.6, and a minimum inclination of about \ang{20} with respect to the binary's orbit. We also find the disc material to have highly eccentric orbits beyond the transition radius, where the disc changes from being dominated by viscous forces, to heavily controlled by the companion star, in contrast to its nearly circular motion inward of the transition radius. Finally, we offer predictions to how these changes will affect Be star observables.
\end{abstract}




\section{Introduction}
\label{sec:intro}

Classical B-emission (Be) stars are defined as rapidly rotating, non-supergiant B-type stars, that have, or have had, Balmer lines in emission \citep{Collins1987}. These emission lines are known to originate from a slowly outflowing circumstellar disc of gas that has formed via some mass-loss mechanism around the equator of the star. Be stars average 70-80\% of their critical rotation velocity \citep{Porter2003}, which is a key influence on the mass-loss efficiency, however the exact mechanisms remain uncertain. It is thought that non-radial pulsations \citep{Baade2016}, and the effect of binary companions \citep{Kriz1975}, may also combine with rapid rotation to pull material off of the star to form a disc. In addition to line emission, Be star discs are also characterised by excess continuum emission \citep{Ghoreyshi2018}, and linear polarization \citep{Halonen2013}.

Currently, the most accepted interpretation of Be star discs is the viscous decretion disc (VDD) model developed by \cite{Lee1991}, which assumes that material is ejected from the star through some unknown mechanism, and then transported outwards through viscosity effects. The VDD model has been successfully implemented in many studies of Be stars \cite[see][for two recent examples]{Ghoreyshi2021, Marr2021}. It has been shown through one-dimensional dynamical modelling that the growth and dissipation of the Be star disc can both brighten and diminish the photometric magnitudes of Be star systems, with corresponding changes in colour, depending on the disc inclination to the observer \citep{Haubois2012, Ghoreyshi2018}, as well as cause characteristic loops in plots of linear polarization in the V-band vs. the polarization at the Balmer Jump \citep{Haubois2014}.

Many Be stars are known to exist in binary systems, and  triple systems involving Be stars have also been known to occur \citep{Rivinius2020, Klement2021}. A survey by \cite{Oudmaijer2010} confirmed 30\% of observed Be stars to have binary companions, the same occurrence rate as their observed normal B-type stars. It was first posed by \cite{Kriz1975} that all Be stars may have binary companions. This idea has gained more support recently from \cite{Klement2019}, who suggested turndown in the radio portion of the spectral energy distribution (SED) of Be stars indicated disc truncation by an undetected binary companion, and by \cite{Bodensteiner2020}, who propose that a lack of main-sequence companion stars in Be binary systems suggests that binary mass transfer is a common disc formation pathway for Be stars. Most recently, \cite{Hastings2021} suggest that, at most, binary interactions can account for one-third of all main-sequence stars being Be stars, and that binary systems can in theory match the observed numbers of Be stars in open clusters.

Many studies have examined the effect that a binary companion has on the structure of a Be star disc. \cite{Okazaki2002} used a 3-dimensional (3D) smoothed particle hydrodynamics (SPH) code to simulate the interaction of a decretion disc of a Be star and a neutron star in a coplanar Be/X-ray binary system. They found significant truncation of the disc due to the binary companion and that the eccentric orbit of the companion induces a two-armed spiral density enhancement at its closest approach. This study was followed up by \cite{panoglou2016discs} and \cite{Cyr2017} who simulated Be binary systems with coplanar and misaligned binary orbits, respectively, while varying the orbital parameters of the secondary. For coplanar orbits, \cite{panoglou2016discs} found that the spiral density enhancements produced by the companion are tidally locked to the binary's orbital phase. They also find the disc appears more truncated and more elongated for a small viscosity. In the misaligned case, \cite{Cyr2017} find a larger truncation radius for a larger misalignment angle between the binary and disc, and also that the misaligned companion can significantly tilt and warp the disc over time. Coplanar simulations were then used by \cite{panoglou2018discs} to show that the spiral enhancements produced by a coplanar binary companion can produce large violet-to-red (V/R) variations in the \Halpha line, as well as triple-peaked and flat-topped emission lines.

The study of misaligned binary companions to Be stars has largely been confined to Be/X-ray systems, where the companion is much less massive than the primary star. In their analytical approach, \cite{Martin2011} find the disc to warp, precess and reach a steady state in timescales of a year to a few hundred years. Through SPH simulations, \cite{Martin2014b} find a high misalignment angle is needed for Type II outbursts to occur in a Be/X-ray binary. \cite{Brown2019} also simulated a variety of Be/X-ray binary systems with an SPH code. They found the disc size was larger with an increased misalignment angle and viscosity, however the Be star disc size does not depend on the mass-ejection rate of the star, all of which are consistent with previously mentioned studies on Be star binary systems. Through analytical investigations involving Lindblad torques, \cite{Miranda2015} and \cite{Lubow2015} have also found that a misaligned disc in a binary system should be more radially extended compared to their coplanar analogue.

One secular effect that may appear in binary Be star systems is Kozai-Lidov (KL) oscillations, which is the exchange of orbital eccentricity and inclination of the inner orbiting body in a 3-body system, first uncovered by \cite{Lidov1962} and \cite{Kozai1962}. KL oscillations were shown by \cite{Martin2014} to be able to occur in hydrodynamical accretion discs in binary star systems, where the disc plays the role of the inner orbiting body. This study was followed up by \cite{Fu2015a} and \cite{Fu2015b}, who showed that KL oscillations can occur in discs for a wide variety of parameters. However they also showed that for a disc of sufficient mass, including the disc self-gravity greatly hindered the occurrence of KL oscillations. Analytical investigations by \cite{Lubow2017} and \cite{Zanazzi2017} have shown KL oscillations are possible in discs given certain conditions for their aspect ratio and sound speed, which are in agreement with discs of the aforementioned SPH studies.

In all previous works of Be stars in binary systems, emphasis has been placed on cases where the binary is coplanar with the disc, and where the disc has already reached a steady configuration. While disc growth has been studied in past works, following both the growth and dissipation of Be star discs has not been done in previous works involving SPH simulations. This paper is the first of two in an effort to examine the observational effects that a misaligned companion can produce in a Be star system as the disc grows and dissipates, in addition to a steady-state phase. It is the goal of this first paper to use SPH simulations to study the evolution and structure of Be star discs in misaligned binary systems through their growth and dissipation phases, as well as their steady state phase. In paper two, we will produce simulated observations of these resulting disc configurations using the radiative transfer code \text{\sc hdust} \citep{carciofi2006non}, to quantify the observational changes that can be seen from the changing disc configurations presented here. Section \ref{sec:Methodology} details the parameters of our simulations, Section \ref{sec:Results} presents the outcomes of these simulations, detailing the disc evolution as well as its steady-state configuration, and in Section \ref{sec:Discussion} we discuss these results, comparing them to other works and implications for observables.

\section{Methodology}
\label{sec:Methodology}

\subsection{Simulation Details}
\label{sec:sim_details}

In this work, we use the same 3D SPH code as past studies by \cite{Okazaki2002}, \cite{panoglou2016discs}, and \cite{Cyr2017}. This code was developed by \cite{Benz1990Book} and \cite{Benz1990}, and later refined by \cite{Bate1995} to include a second-order Runge-Kutta-Fehlberg integrator. It was then adapted by \cite{Okazaki2002} to simulate decretion discs around Be stars.

The geometry of the system is as follows: the spin axis of the primary star points along the $z$-axis, and hence the primary's equatorial plane is the $x-y$ plane. This $x-y$ plane is also the plane where the particles are injected from the primary star into the disc. The misalignment angle of the binary companion is measured from the $x$-axis. The line of nodes is coincidental with the $y$-axis

Our simulations involve an equal-mass binary system in a circular orbit, with orbital periods of either 30 or 300 days, in which both the central and secondary star are programmed as sink particles. These sink particles have defined accretion radii inside which any particles are assumed to be accreted and are removed from the simulation. The accretion radius for the primary star is equal to its radius, while the accretion radius of the secondary star is equal to the size of its Roche lobe. Using the approximation of \cite{Eggleton1983}, this Roche lobe radius is about $0.38a$, where $a$ is the distance between the two sink particles. Due to the circular orbits in our simulations, this Roche lobe radius is constant. The binary orbit is misaligned from the equatorial plane of the primary star by either \ang{20}, \ang{40}, or \ang{60}. The SPH code utilizes the Shakura-Sunyaev viscosity prescription where the viscosity, $\nu$, is defined as \citep{Shakura1973}
\begin{equation}
    \nu = \alpha c_s H,
\end{equation}
where $c_s$ is the disc sound speed, $H$ is the disc scale height, and $\alpha$ is a dimensionless scaling parameter. In all of our simulations, we set $\alpha\, = 0.1$. While $\alpha$ has been shown to range from 0.1 - 1 for Be stars \citep{Rimulo2018}, recent works indicate $\alpha$ to be towards the lower end of this interval \citep{Ghoreyshi2021, Marr2021}. The value of 0.1 is also frequently used in SPH studies previously mentioned, so using 0.1 here facilitates easy comparison to past works.

At the beginning of the simulation there is no disc around the primary star, and at each subsequent timestep of size $1/200\pi$ orbital periods, 5000 equal-mass particles are injected into the simulation at the injection radius, $r_{\rm inj}\, =\, 1.04 \, \rm R_p$ \cite[same as][]{Cyr2017}, in the equatorial plane of the primary star. The particle mass is calculated based on our chosen mass-injection rate, $\dot{M}_{inj}\, =\, 10^{-8} \, \rm M_\odot yr^{-1}$ from the primary star into the disc. Note that most of the injected mass falls back onto the star almost immediately and only a small fraction of this actually stays in the disc \cite[see fig. 5 of][]{Okazaki2002}. To simulate the growth and dissipation of the disc, we keep this mass-injection turned on for 100 orbital periods (\Porb), then turn it off at 100 \Porb \text{ }and allow the disc to dissipate for the following 100 \Porb, or until there is an insufficient number of particles for the simulation to continue. Our simulation parameters are summarized in Table \ref{tab:sph_sim_params}.

\begin{table}
\centering
\caption{Simulation parameters of our SPH models. Values for the primary and secondary star have respective subscripts $p$ and $s$.}
\label{tab:sph_sim_params}
\begin{threeparttable}[c]
\renewcommand{\TPTminimum}{\linewidth}
\makebox[\linewidth]{
\vspace{3mm}
\begin{tabular}[c]{cc}
\hline\hline
    Parameter & Value  \\
    \hline
    $M_{\rm p}$ & 8 $\rm M_\odot$ \\
    $R_{\rm p}$ & 5 $\rm R_\odot$ \\
    $T_{\rm p}$ & 20000 K \\
    $M_{\rm s}$ & 8 $\rm M_\odot$ \\
    $R_{\rm s}$ & 5 $\rm R_\odot$ \\
    $T_{\rm disc}$ & 12000 K \tnote{a} \\
    $\alpha$ & 0.1 \\ 
    $\dot{M}_{\rm inj}$ & $10^{-8} \, \rm M_\odot yr^{-1}$ \\
    $r_{\rm inj}$ & $1.04 \, \rm R_p$ \\
    Misalignment Angle & \ang{20}/\ang{40}/\ang{60} \\ 
    Orbital Period & 30/300 days\\
    Orbital Radius & 20.5/95 $\rm R_p$ \\
\hline
\end{tabular}}
\begin{tablenotes}
\item[a] The disc temperature is set to 60\% of the primary star's effective temperature. This value was found by \cite{carciofi2006non} to be the average temperature in the isothermal regions of the disc.
\end{tablenotes}
\end{threeparttable}
\end{table}

The parameters chosen here of course represent a small subset of the possible combinations, however they are in line with past SPH simulations of Be star discs \citep{Cyr2017, panoglou2016discs}, as well as accretion discs \citep{Martin2014} in misaligned binary systems. The similarity in parameters allows for smooth comparison of this work to the past mentioned works, while still offering new interesting results. The computational requirements for SPH simulations such as those presented here does not lend itself to efficiently covering a large range of parameters at once, thus our initial choices here also leave the opportunity for future investigations of other parameter combinations.

\subsection{Calculation Details}
\label{sec:calc_details}

In the analysis of the disc evolution from our simulations, we find the disc inclination and eccentricity by calculating these quantities separately for all particles, and then averaging the particle values into one value for the disc. We first determine the specific angular momentum, $\mathbf{j}$, of each particle relative to the primary star, through
\begin{equation}
    \mathbf{j} = (\mathbf{r} - \mathbf{r_p}) \times (\mathbf{v} - \mathbf{v_p}),
    \label{eq:ang_mom}
\end{equation}
where $\mathbf{r}$ and $\mathbf{v}$ are the position and velocity vectors of the particle, and the subscript p denotes the same values but for the primary star. The specific energy of the particle is then computed by 
\begin{equation}
    E = \frac{1}{2}|\mathbf{v} - \mathbf{v_p}|^2 - \frac{GM_p}{|\mathbf{r} - \mathbf{r_p}|},
    \label{eq:spec_energy}
\end{equation}
with $G$ the gravitational constant, and $M_p$ the mass of the primary star. The particle's inclination, $i$, and eccentricity, $e$, are given as
\begin{eqnarray}
    i &=& \arccos{\frac{j_z}{|\mathbf{j}|}},\\
    e &=& \sqrt{1 + \frac{2E|\mathbf{j}|^2}{(GM_p)^2}}.
\end{eqnarray}
The total disc mass and angular momentum are likewise found by simply summing these values for all active particles in the disc.

\section{Results}
\label{sec:Results}

\subsection{Disc Evolution}
\label{Sec:disc_evol}

This section presents the resulting evolution of the disc in our six individual simulations. For brevity, we shall shorten the names of our simulations. For example, our simulation with a 30 day orbital period and \ang{20} misalignment of the binary orbit, will be written as ``our 30 day, \ang{20} simulation" and so forth.

\subsubsection{30 Day, \ang{20} Simulation}
\label{sec:30/20}

\begin{figure}
    \centering
    \includegraphics[scale = 0.4]{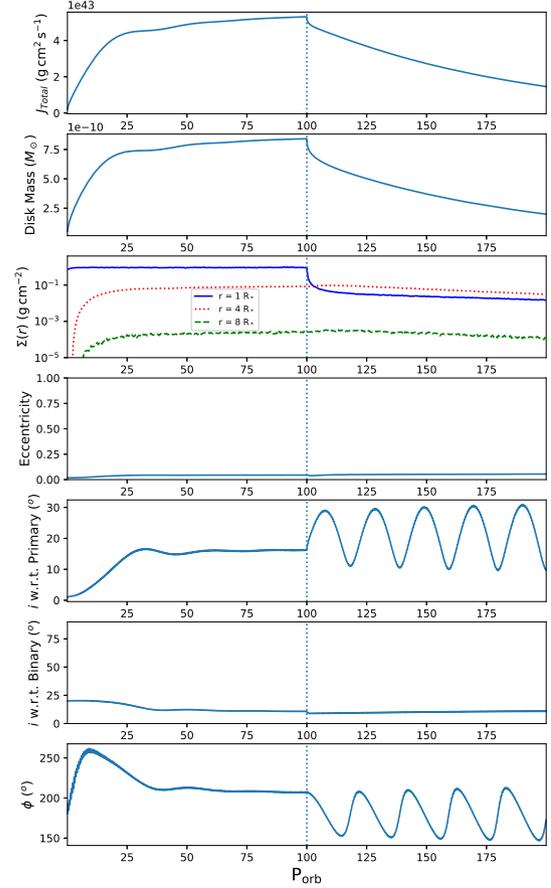}
    \caption{Top to bottom, evolution of the disc angular momentum, mass, azimuthally averaged surface density at 1 $R_*$ (blue, solid), 4 $R_*$ (red, dotted), and 8 $R_*$ (green, dashed), average eccentricity, inclination with respect to the primary's equatorial plane, inclination with respect to the binary's orbital plane, and the longitude of the ascending node of the disc measured from the positive $x$-axis for the 30 day, \ang{20} simulation. The $x$-axis is in units of binary orbital periods. The vertical dotted line indicates the point where the mass-injection was turned off.}
    \label{fig:30/20_total_ang_mom}
\end{figure}

The evolution of the disc in our 30 day, \ang{20} simulation is shown in Figure \ref{fig:30/20_total_ang_mom} where we present the disc angular momentum, mass, azimuthally averaged surface density, eccentricity, inclination with respect to the primary's equatorial plane, inclination with respect to the binary's orbital plane, and the longitude of the ascending node of the disc of over the whole simulation. The longitude of the ascending node is calculated from the positive $x$-axis in our simulations, with the reference plane for the line of nodes of the disc being the equatorial plane of the primary star. As seen in the fifth and sixth panel, while the disc is growing the torque from the binary companion tilts the disc towards the binary's orbital plane and the two nearly align before the disc settles at an inclination of about \ang{15} after 60  \Porb. The disc does not fully align with the binary's orbital plane due to the constant mass-injection at the equator of the primary star, with the innermost disc, while technically  free to move, being bound by gravity to this equatorial plane. The third panel supports this by showing that the inclination of the disc does not increase dramatically until the outer disc reaches a fairly steady surface density and has a higher number of particles, which would be most tilted by the binary companion. 

When disc dissipation begins at 100 \Porb \text{ }the disc mass initially drops sharply as the inner disc almost immediately reaccretes, and then drops steadily afterwards as the rest of the disc gradually falls back onto the primary star. Throughout the simulation, the disc's average eccentricity is negligible ($e < 0.055$), however the inclination of the disc oscillates greatly once dissipation begins with an amplitude of almost \ang{10}. This oscillation is due to the disc precessing about the binary's orbital axis, which can be seen in the last panel of Figure \ref{fig:30/20_total_ang_mom}, where the longitude of the ascending node oscillates between 150 and 200 degrees. By taking the Fourier transform of the x-component of angular momentum, we find the precession period here to be 20 \Porb.


\subsubsection{30 Day, \ang{40} Simulation}
\label{sec:30/40}

\begin{figure}
    \centering
    \includegraphics[scale = 0.4]{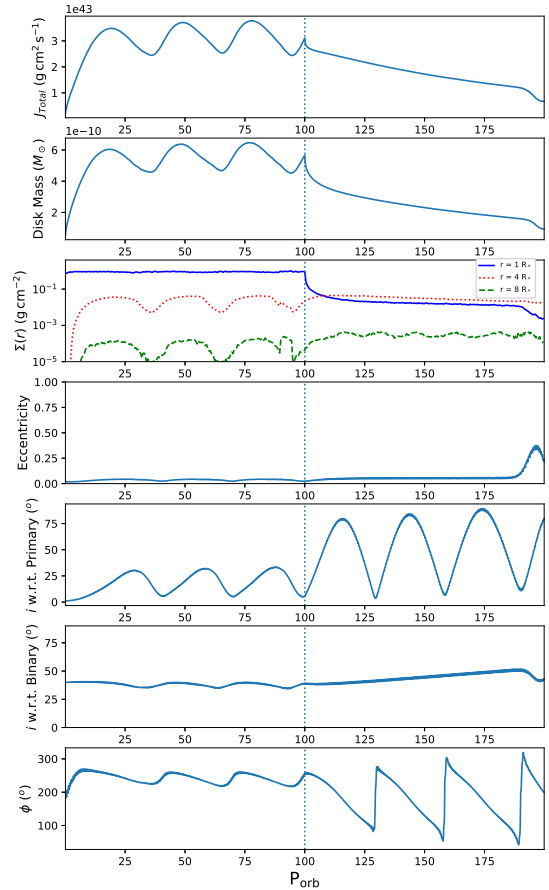}
    \caption{Same as Figure \ref{fig:30/20_total_ang_mom}, but for the 30 day, \ang{40} simulation.}
    \label{fig:30/40_total_ang_mom}
\end{figure}

\begin{figure*}
    \centering
    \includegraphics[scale = 0.3]{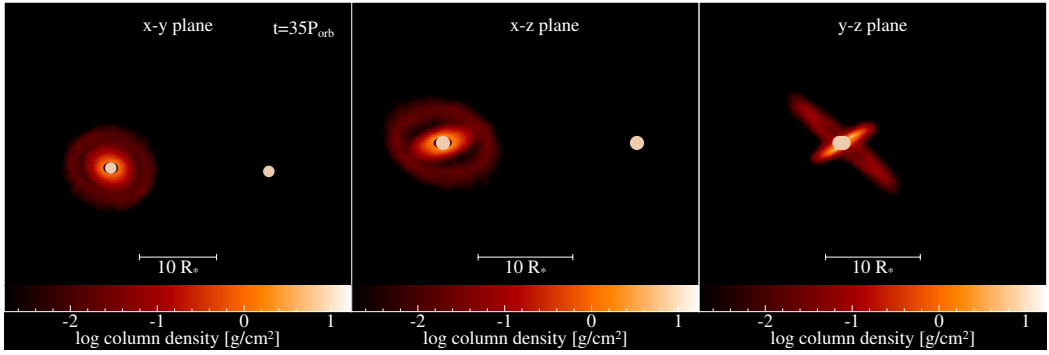}
    \caption{Snapshot of our 30 day, \ang{40} simulation at 35 \Porb. Left to right shows the $x-y$, $x-z$, and $y-z$ planes. The primary and secondary star are represented by white circles, and the disc is coloured by its column density, indicated by the colour bars under each window. The scale bar in each window indicates the length of 10 primary stellar radii ($R_*$).}
    \label{fig:30/40_35Porb}
\end{figure*}

The 30 day, \ang{40} simulation starts very similarly to the \ang{20} case, as seen in Figure \ref{fig:30/40_total_ang_mom}, with the disc moving towards aligning with the binary companion. However in this case, at around 25 to 30 \Porb, the disc tears into two separate pieces, creating a separate tilted ring of material around the remaining disc. This ring then precesses independently before recombining with the inner portion of the disc. This behaviour continues periodically on approximately 30 \Porb \text{ }periods until the mass-injection into the disc is turned off. The effects of this can be seen in Figure \ref{fig:30/40_total_ang_mom}, where the calculated quantities of the disc all oscillate in phase with these tearing episodes before dissipation. However it can also be seen that only the outer disc is affected, as the third panel of Figure \ref{fig:30/40_total_ang_mom} shows no change to the inner disc surface density prior to dissipation. A visual snapshot of this phenomenon is shown in Figure \ref{fig:30/40_35Porb}.

\begin{figure*}
    \centering
    \includegraphics[scale = 0.3]{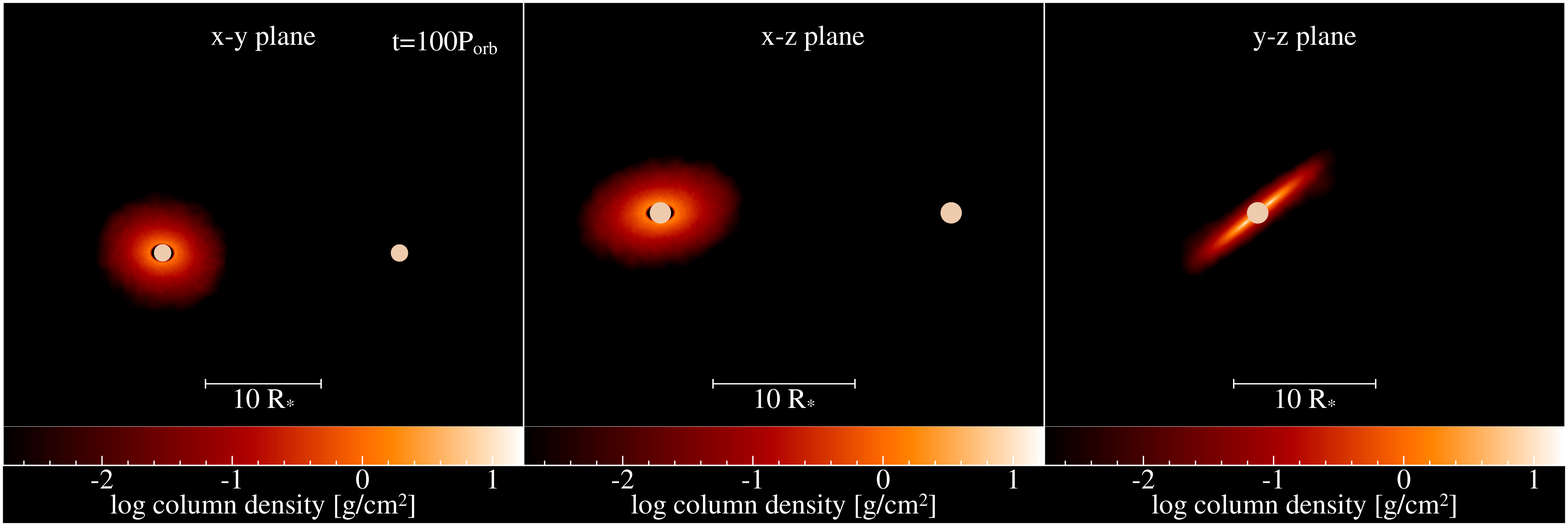}
    \includegraphics[scale = 0.3]{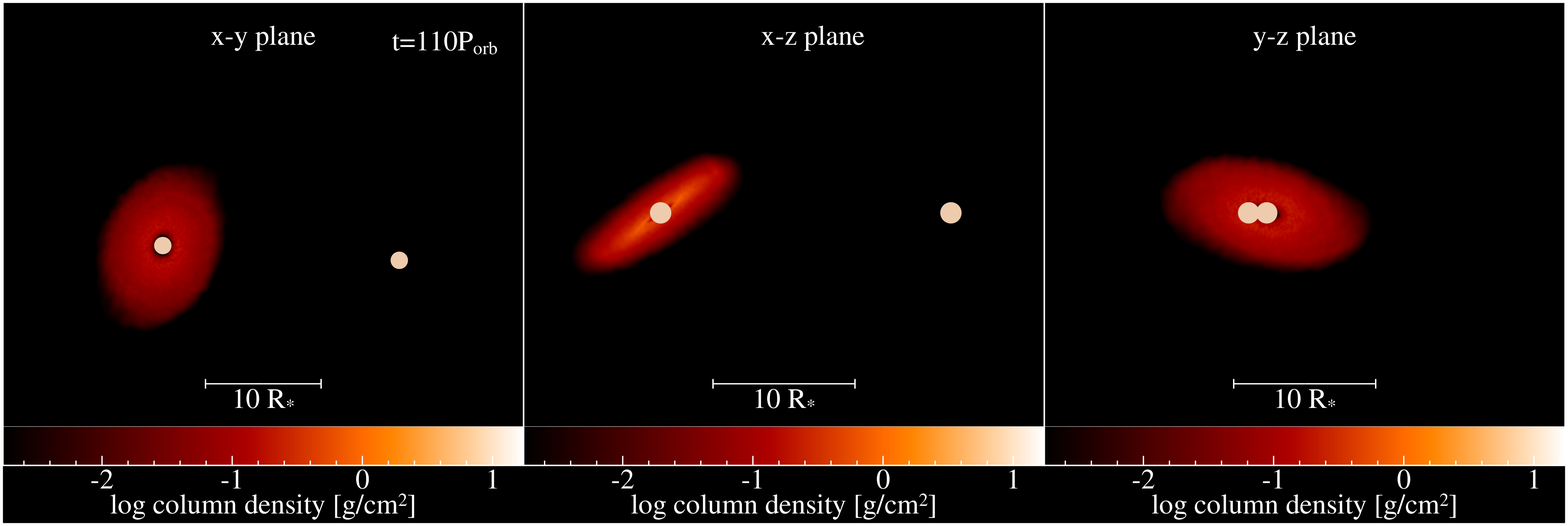}
    \includegraphics[scale = 0.3]{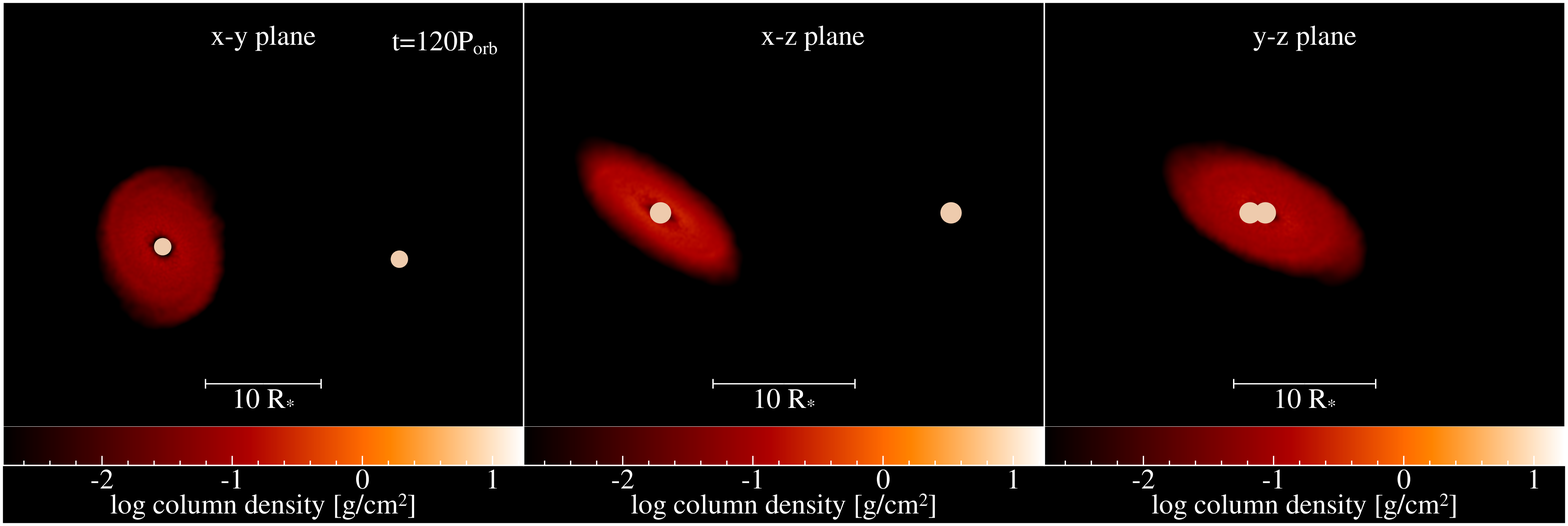}
    \caption{Snapshots of the 30 day, \ang{40} simulation at (top to bottom) 100, 110, and 120 \Porb. Format is the same as Figure \ref{fig:30/40_35Porb}.}
    \label{fig:30/40_100-120Porb}
\end{figure*}


\begin{figure}
    \centering
    \begin{subfigure}[b]{0.39\textwidth}
         \centering
         \includegraphics[width=\textwidth]{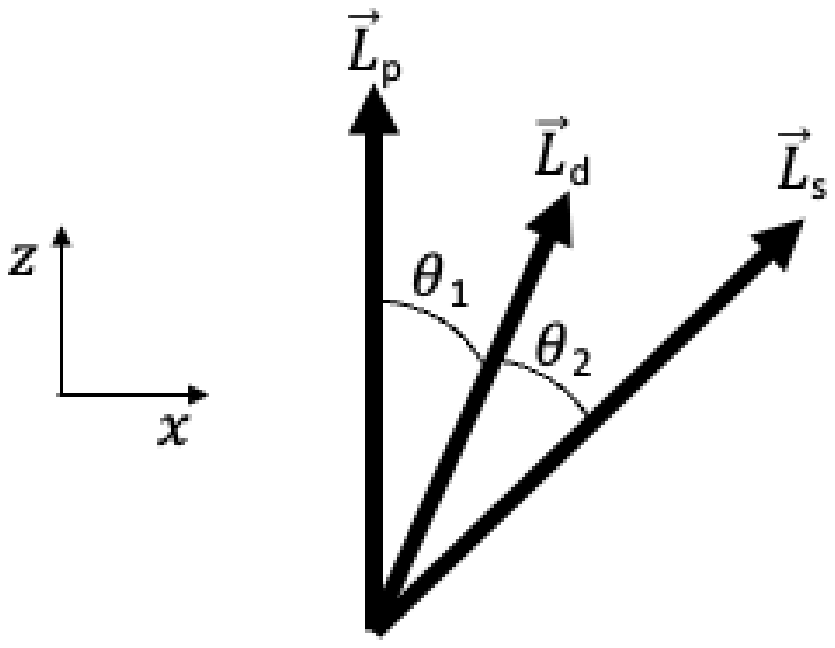}
         \caption{Side View}
         \label{fig:SPH_side_view}
     \end{subfigure}
     \begin{subfigure}[b]{0.39\textwidth}
         \centering
         \includegraphics[width=\textwidth]{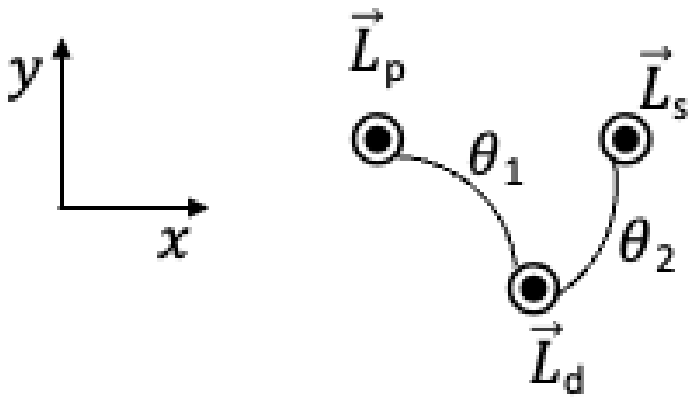}
         \caption{Top View}
         \label{fig:SPH_top_view}
     \end{subfigure}
    \caption{Schematic showing the difference between the inclination of the disc with respect to the primary star and the binary companion. The inclinations are respectively labelled $\theta_1$ and $\theta_2$. The angular momentum vectors of the primary star, disc, and binary companion are respectively labelled $L_{\rm p}$, $L_{\rm d}$, and $L_{\rm s}$. (a) shows the side view of the vectors, from the $x-z$ plane, while (b) shows the same vectors from the top, looking at the $x-y$ plane.}
    \label{fig:SPH_schematic}
\end{figure}

When the mass-injection is turned off, coincidentally the disc is not split into two. The disc no longer undergoes this periodic tearing, but rather it precesses as a whole in the same manner as the \ang{20} case, but with a longer precession period of about 33 \Porb. This behaviour is displayed in Figure \ref{fig:30/40_100-120Porb} where we can see the orientation of the disc can change dramatically due to this precession, depending on the point of view of the observer. In the last 10 orbital periods of the simulation, we detect KL oscillations occurring. This can be seen in the fourth, fifth, and sixth panels of Figure \ref{fig:30/40_total_ang_mom} where, near the very end of the simulation, we see the KL oscillations beginning, with the eccentricity and the disc's inclination from the binary's orbital plane interchanging dramatically. Note that, as shown in the schematic diagram in Figure \ref{fig:SPH_schematic}, the two inclinations calculated here do not change with the same amplitude due to the three-dimensional nature of the simulation. While the angular momentum vectors for the primary and secondary star are fixed, the angular momentum vector of the disc can move in any direction, thus the angle with respect to one star could vary largely while the angle with respect to the other star stays constant.

\subsubsection{30 Day, \ang{60} Simulation}
\label{sec:30/60}

\begin{figure}
    \centering
    \includegraphics[scale = 0.4]{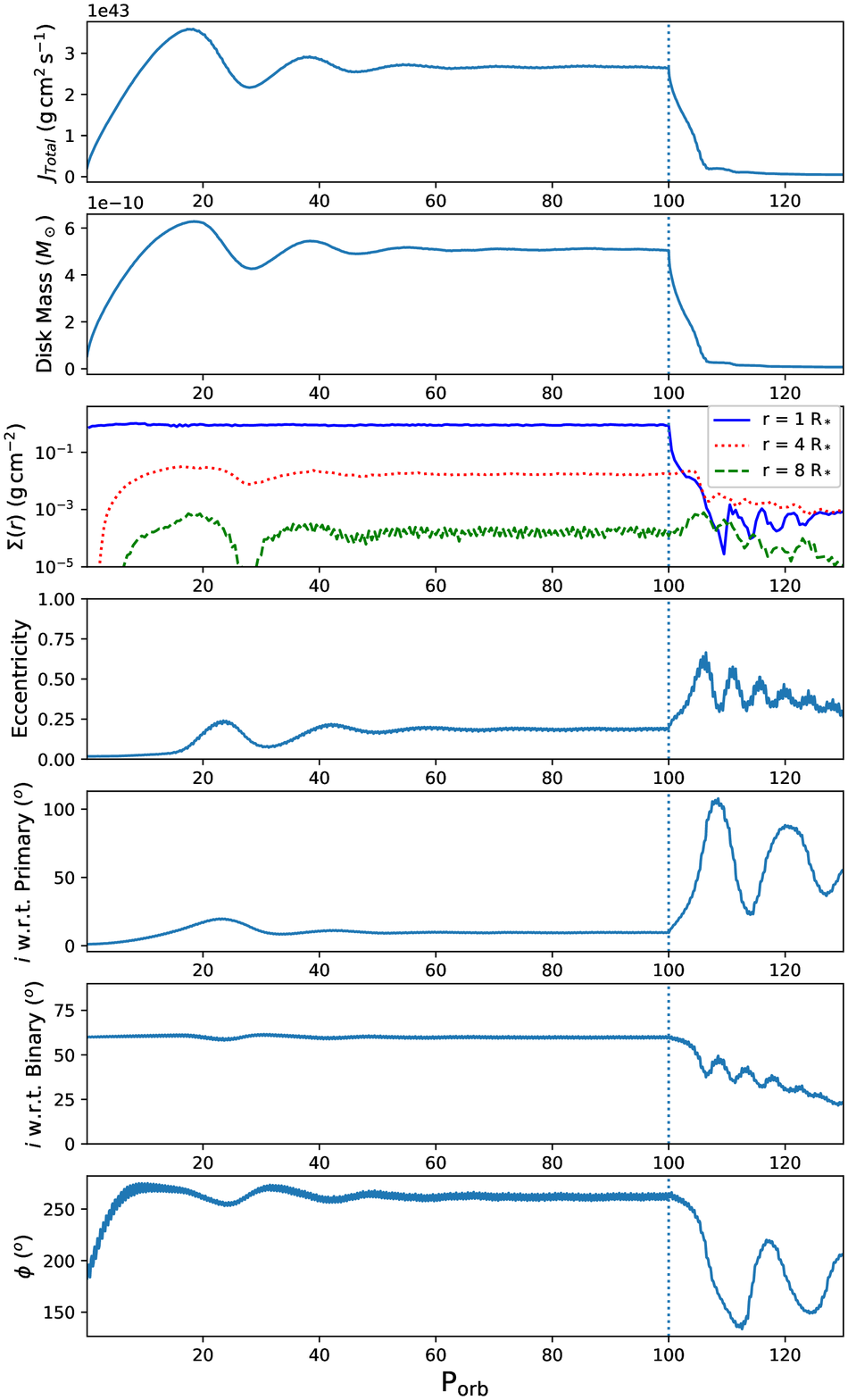}
    \caption{Same as Figure \ref{fig:30/20_total_ang_mom}, but for the 30 day, \ang{60} simulation.}
    \label{fig:30/60_total_ang_mom}
\end{figure}

The evolution of the 30 day, \ang{60} simulation is shown in Figure \ref{fig:30/60_total_ang_mom}. We again see that the disc inclination does not change dramatically until the outer disc is populated, however this simulation shows much different behaviour than the previous \ang{20} and \ang{40} cases, particularly after dissipation begins.

\begin{figure*}
    \centering
    \includegraphics[scale = 0.3]{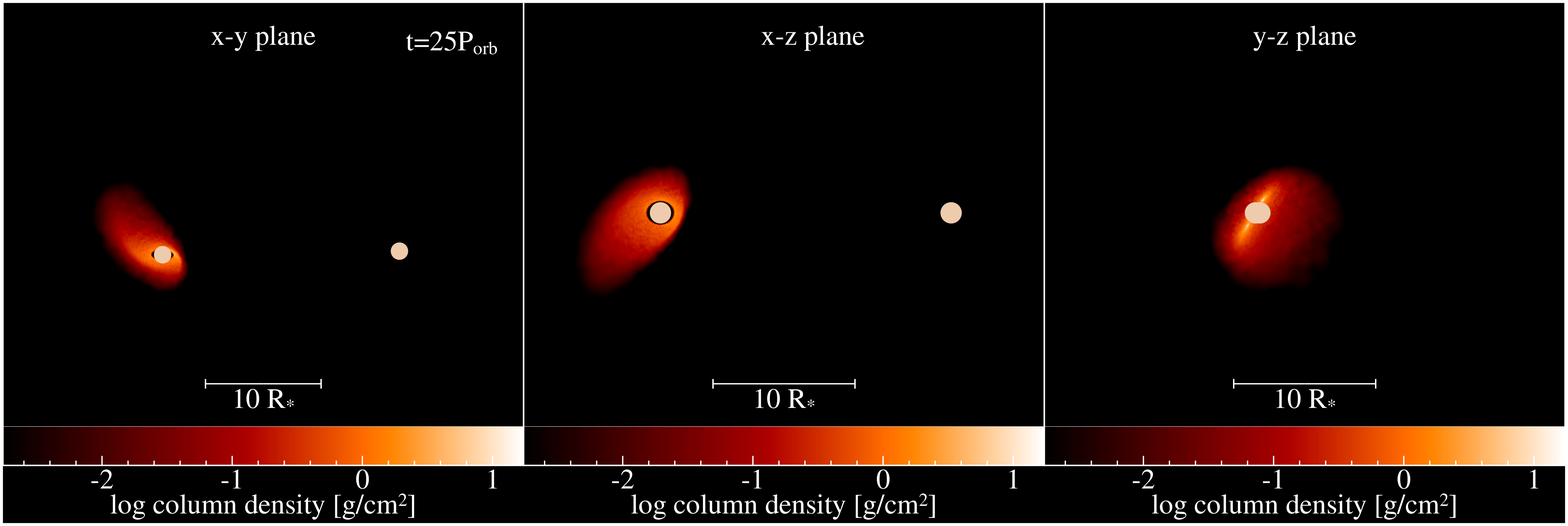}
    \caption{Snapshot of the 30 day, \ang{60} simulation at 25 \Porb. Format is the same as Figure \ref{fig:30/40_35Porb}.}
    \label{fig:30/60_25_Porb}
\end{figure*}

The previous cases displayed negligible average disc eccentricity, however looking at the fourth panel of Figure \ref{fig:30/60_total_ang_mom}, in the \ang{60} case, the disc quickly becomes moderately eccentric due to the effect of the binary companion. Before dissipation we see that the eccentricity and inclination of the disc oscillate before settling to values of 0.2 and \ang{10} respectively, but the inclination with respect to the binary's orbital plane does not change noticeably during this mass-injection phase, which indicates this is not a KL oscillation. A snapshot showing the eccentricity of the disc at 25 \Porb \text{ } is shown in Figure \ref{fig:30/60_25_Porb}.

\begin{figure*}
    \centering
    \includegraphics[scale = 0.3]{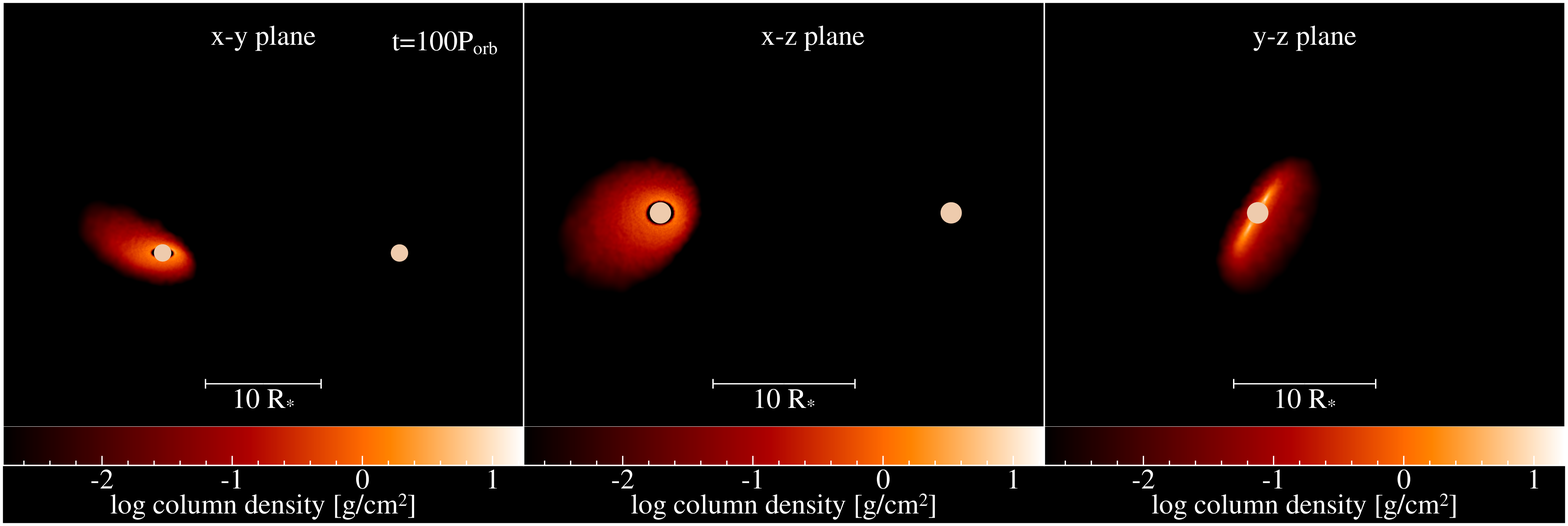}
    \includegraphics[scale = 0.3]{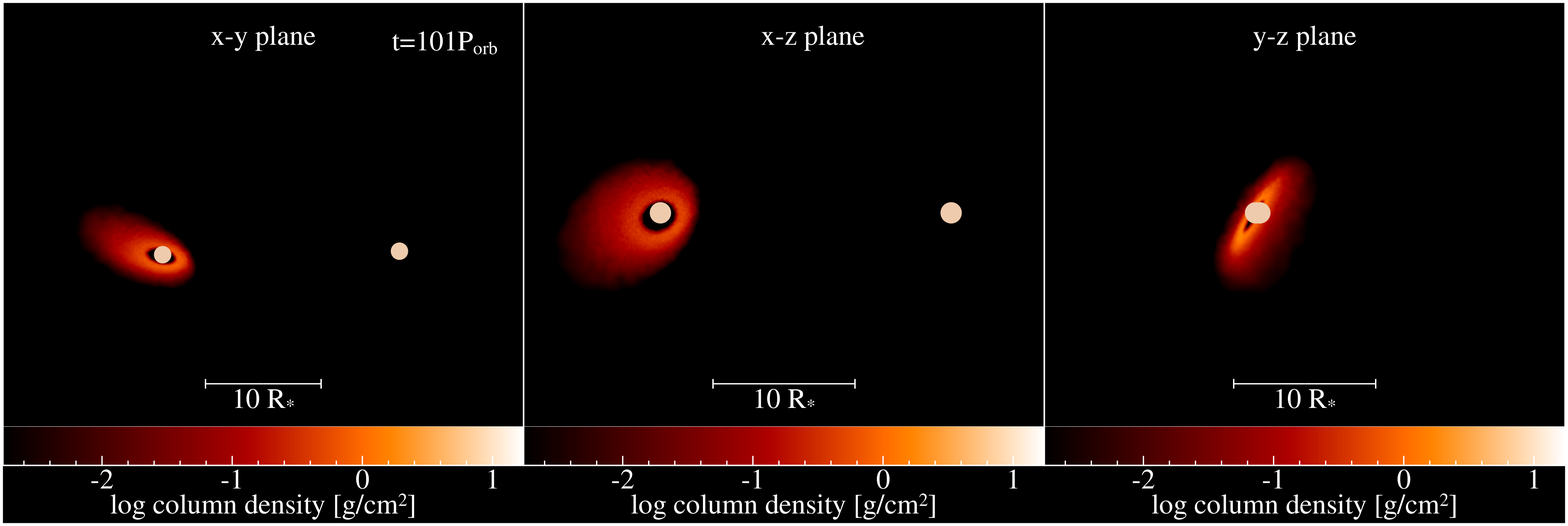}
    \includegraphics[scale = 0.3]{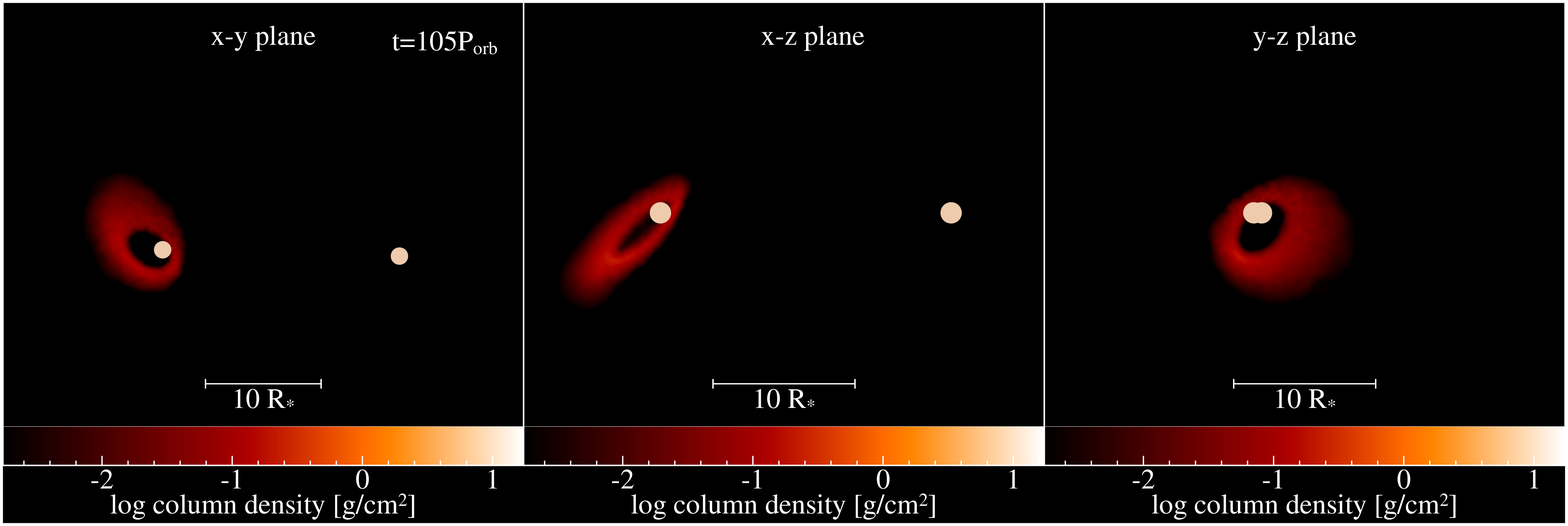}
    \includegraphics[scale = 0.3]{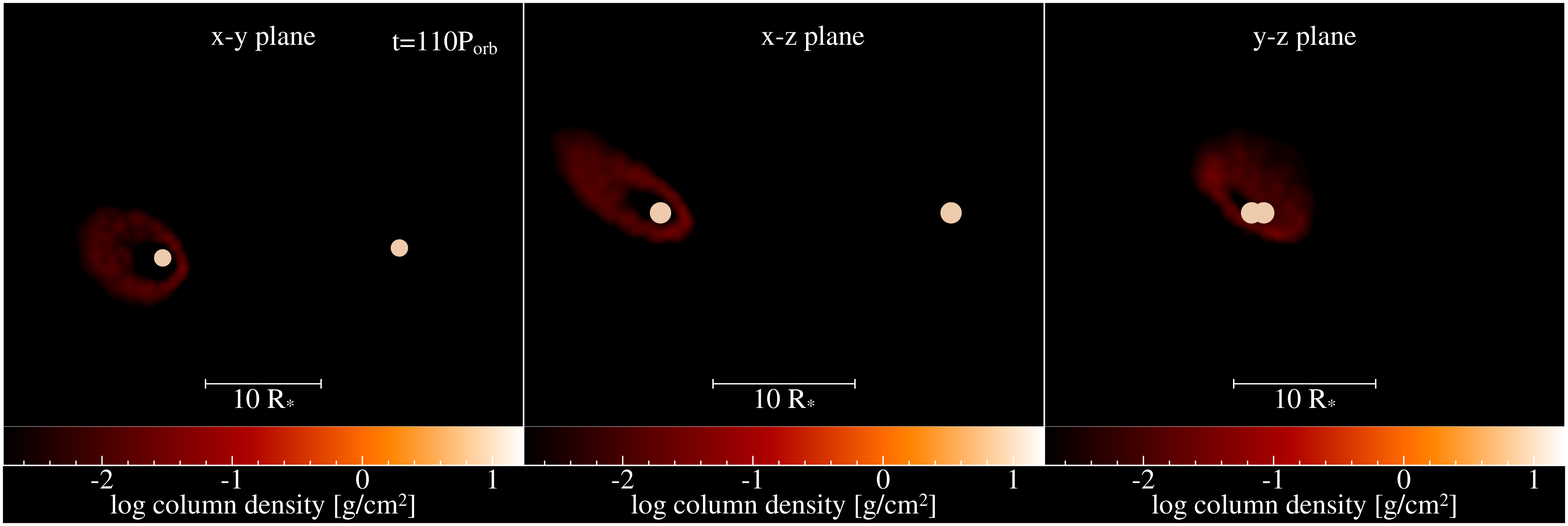}
    \caption{Snapshots of the 30 day, \ang{60} simulation at (top to bottom) 100, 101, 105, and 110 \Porb. Format is the same as Figure \ref{fig:30/40_35Porb}.}
    \label{fig:30/60_100-110Porb}
\end{figure*}

Once disc dissipation begins, the disc undergoes KL oscillations, exchanging eccentricity and inclination throughout the dissipation phase. We also see the inclination of the disc with respect to the primary's equatorial plane grow to greater than \ang{90}, indicating the disc has crossed over the pole of the primary star, and is now orbiting in retrograde fashion. As well, due to the initial eccentricity of the disc at the onset of dissipation, the disc dissipates unevenly and an eccentric gap develops between the primary star and disc. These effects are displayed in the snapshots of the disc seen in Figure \ref{fig:30/60_100-110Porb}.

The analytical timescale for KL oscillations in the case of a test particle in an initially circular orbit is given by \citep{Kiseleva1998}
\begin{equation}
    \frac{\tau_{\rm KL}}{P_b} \approx \frac{M_1 + M_2}{M_2}\frac{P_b}{P_p}(1-e_b^2)^{3/2},
    \label{eq:KL_simple}
\end{equation}
where $M_1$ and $M_2$ are respectively the primary and secondary star masses, $P_b$ is the binary orbital period, $P_p$ is the particle's orbital period about the primary star, and $e_b$ is the binary orbital eccentricity. \cite{Martin2014} also present an estimate for a global disc response period as
\begin{equation}
    \langle \tau_{\rm KL} \rangle \approx \frac{\int_{R_{in}}^{R_{out}}\Sigma R^3 \sqrt{\frac{G M_1}{R^3}}dR}{\int_{R_{in}}^{R_{out}}\tau_{KL}^{-1}\Sigma R^3 \sqrt{\frac{G M_1}{R^3}}dR},
    \label{eq:KL_complex}
\end{equation}
with $\Sigma$ being the disc surface density, $R$ the radial position in the disc, $G$ the gravitational constant, and the integral being computed over the whole disc. Using the parameters of our simulation and the azimuthally averaged surface density at the start of disc dissipation, we calculate the KL timescales should be 2.7 \Porb \text{ }from Equation \ref{eq:KL_simple}, and 33 \Porb \text{ }according to Equation \ref{eq:KL_complex}. \cite{Martin2019} find that a test particle with an initial eccentricity of 0.2 has a KL timescale 2.7 times shorter than what is analytically predicted. This would bring the prediction of Equation \ref{eq:KL_complex} to 12.2 \Porb. The first peak in the disc eccentricity in Figure \ref{fig:30/60_total_ang_mom} occurs after about 6 \Porb, resulting in a KL timescale around 12 \Porb \text{ }during the initial dissipation, in agreement with the analytical prediction.

\subsubsection{300 Day Simulations}
\label{sec:300_day}

\begin{figure*}
    \centering
    \begin{subfigure}[b]{0.35\textwidth}
         \centering
         \includegraphics[width=\textwidth]{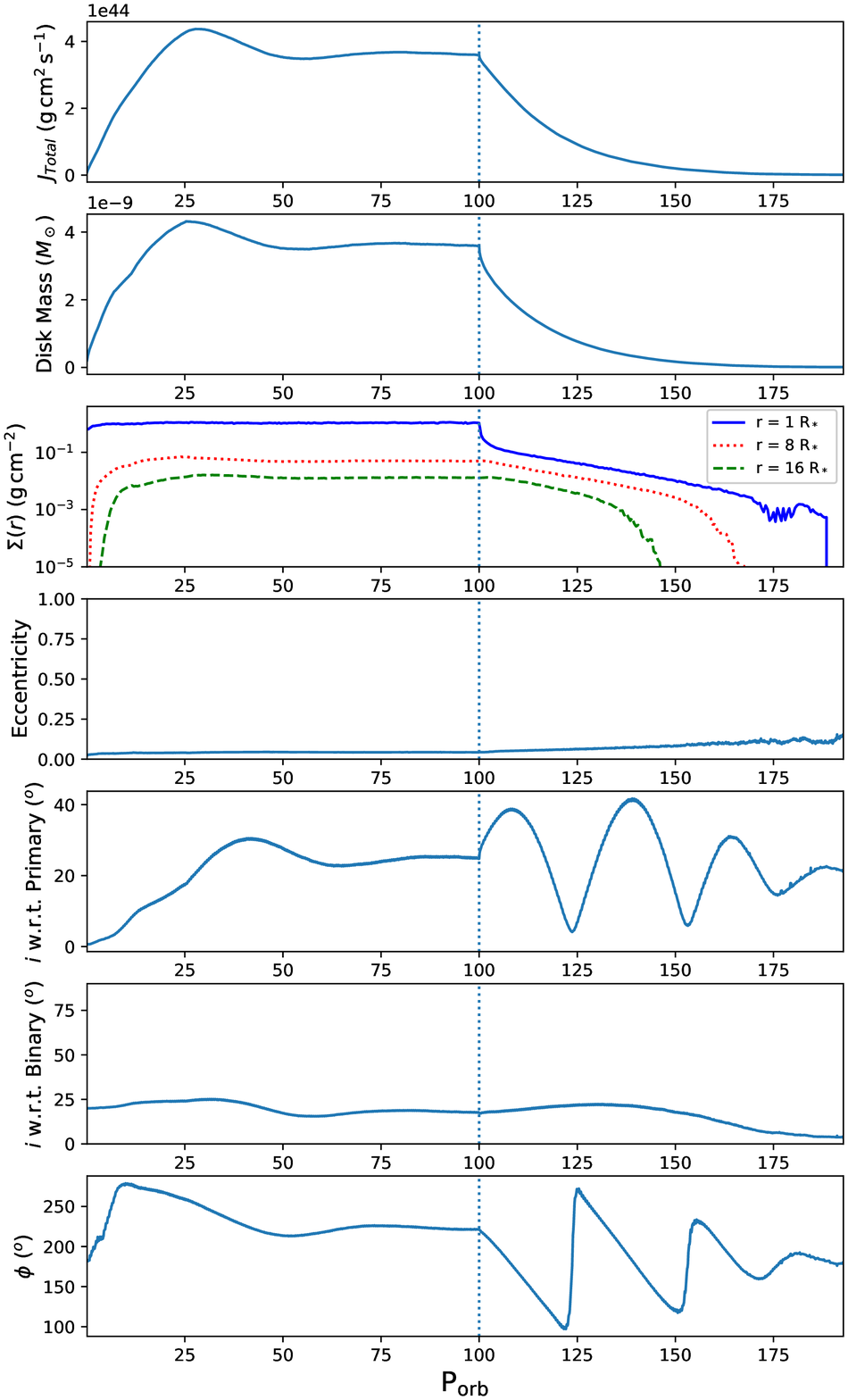}
         \caption{300 day, \ang{20}}
         \label{fig:300/20_tot_ang_mom}
     \end{subfigure}
     \begin{subfigure}[b]{0.35\textwidth}
         \centering
         \includegraphics[width=\textwidth]{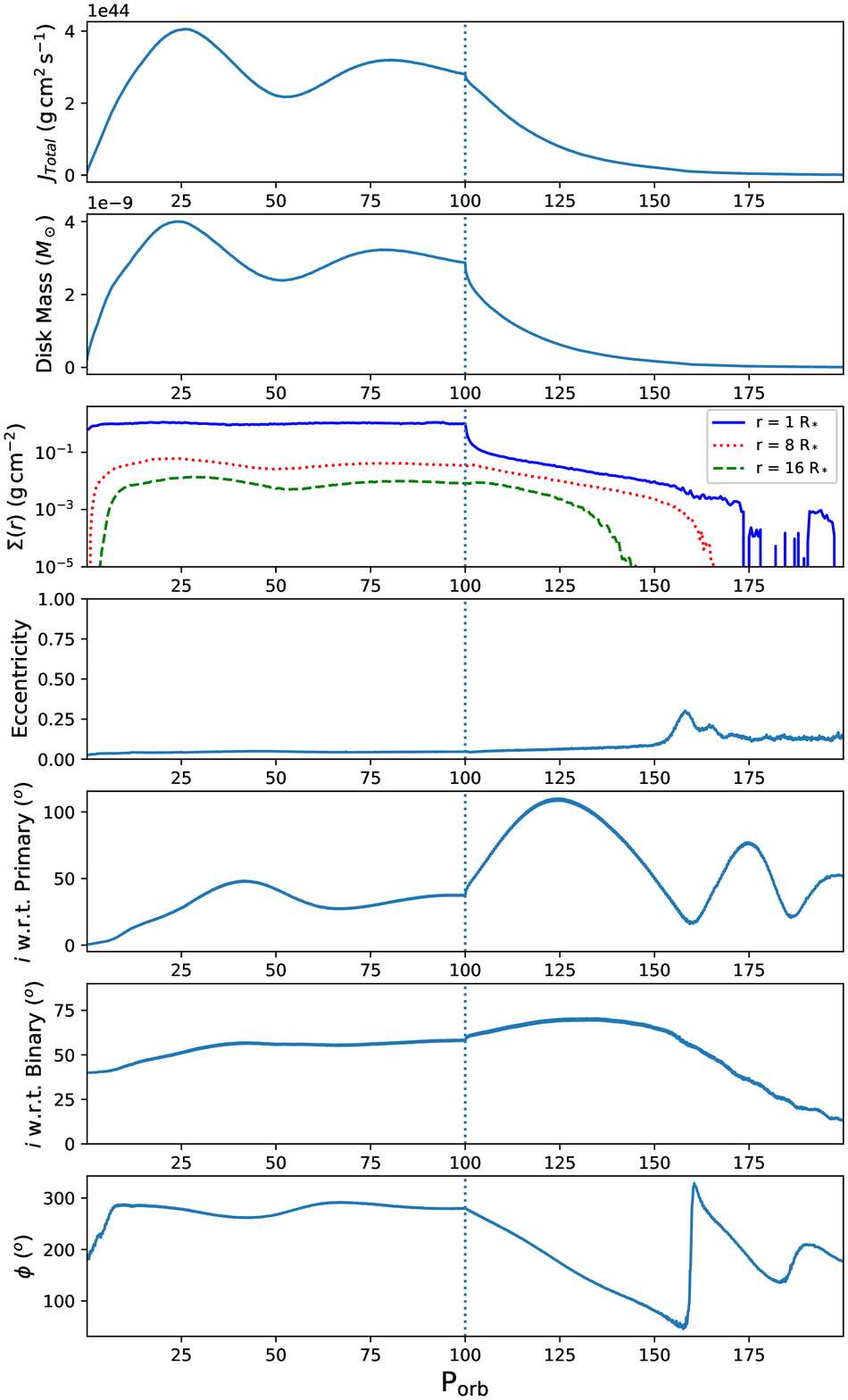}
         \caption{300 day, \ang{40}}
         \label{fig:300/40_tot_ang_mom}
     \end{subfigure}
     \begin{subfigure}[b]{0.35\textwidth}
         \centering
         \includegraphics[width=\textwidth]{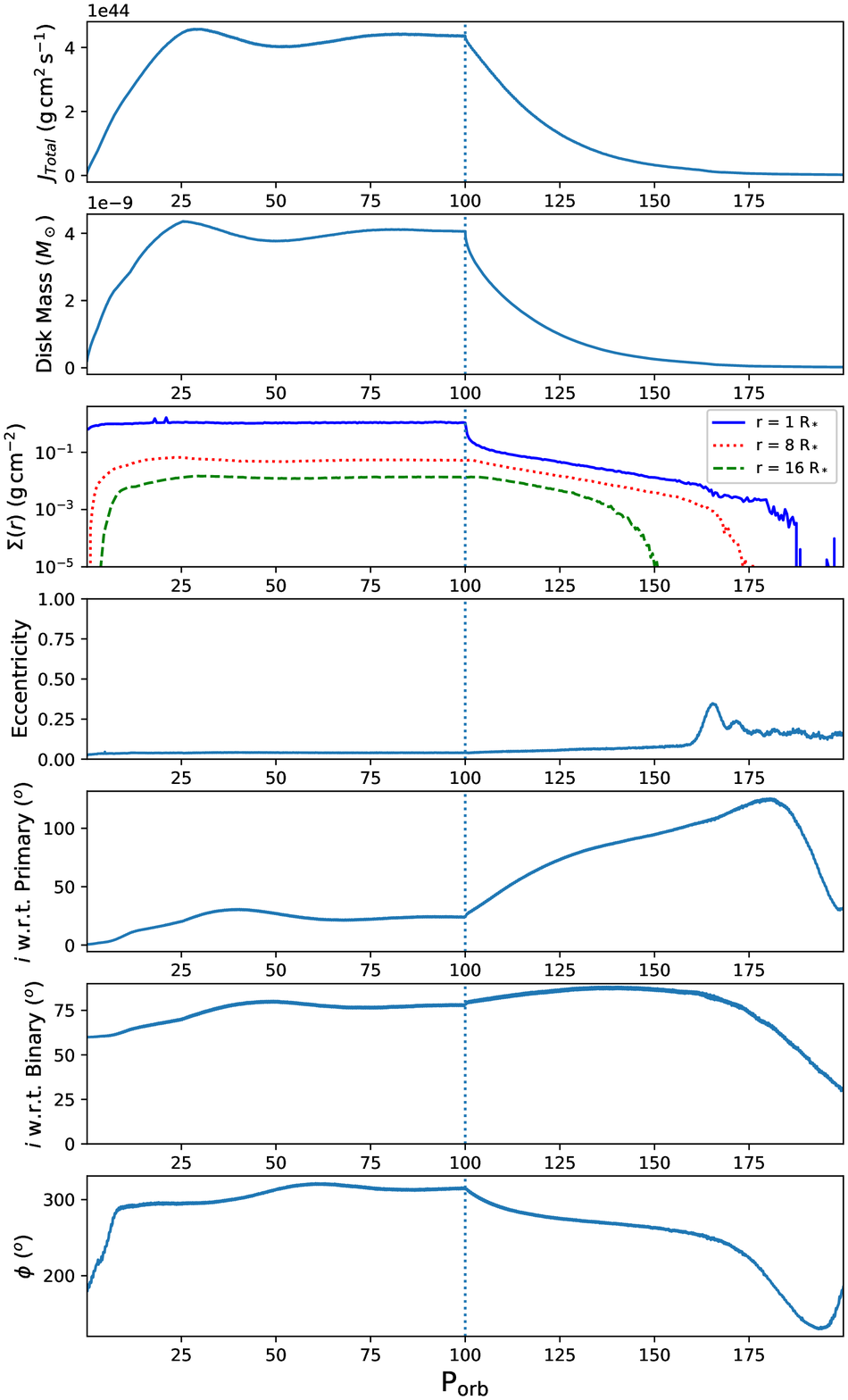}
         \caption{300 day, \ang{60}}
         \label{fig:300/60_tot_ang_mom}
     \end{subfigure}
    \caption{Same format as Figure \ref{fig:30/20_total_ang_mom}, but for the 300 day simulations. The surface densities are calculated at 1 $R_*$ (blue, solid), 8 $R_*$ (red, dotted), and 16 $R_*$ (green, dashed).}
    \label{fig:300_total_ang_mom}
\end{figure*}

Figure \ref{fig:300_total_ang_mom} shows the evolution of the disc in each of the three 300 day simulations. All of the 300 day simulations follow essentially the same evolution pattern amongst themselves: the disc is tilted some amount by the binary companion during the mass injection phase, and then precesses about the binary orbital axis once the mass-injection is turned off and the disc dissipation begins. None of the 300 day simulations show any considerable eccentricity throughout the simulation. During their precession, the \ang{40} and \ang{60} simulations also have a period of time where the discs orbit the primary star in a retrograde manner (inclination > \ang{90}). Note that, for the 300 day, \ang{60} simulation, the disc dissipates before one precession period can complete as seen in the fifth panel of Figure \ref{fig:300/60_tot_ang_mom}. In the case of the \ang{40} and \ang{60} misalignments, the disc tilts away from the binary's orbital plane, making the misalignment angle greater, during the mass-injection phase. This behaviour is interpreted as tilt instabilities arising from the tidal effects of the binary companion, described by \cite{Lubow1992} and also seen, for example, in \cite{Martin2020}. This is in contrast to all other simulations, both with 30 and 300 day orbital periods, in which the disc tilts toward alignment with the binary's orbital plane.

Another commonality in all simulations, including the previously discussed short-period simulations (except for the 30 day, \ang{20} case), is that the tilting of the disc during the mass-injection phase can cause some of the disc to dissipate prematurely back onto the primary star, and the overall disc mass can drop while the mass-injection is still on. This is seen in Figures \ref{fig:30/40_total_ang_mom}, \ref{fig:30/60_total_ang_mom}, and \ref{fig:300_total_ang_mom}, where we can see that the disc angular momentum and mass oscillate during the mass-injection phase in the same manner as the inclination, though they both consistently peak slightly earlier than the latter.

\begin{figure*}
    \centering
    \includegraphics[scale = 0.3]{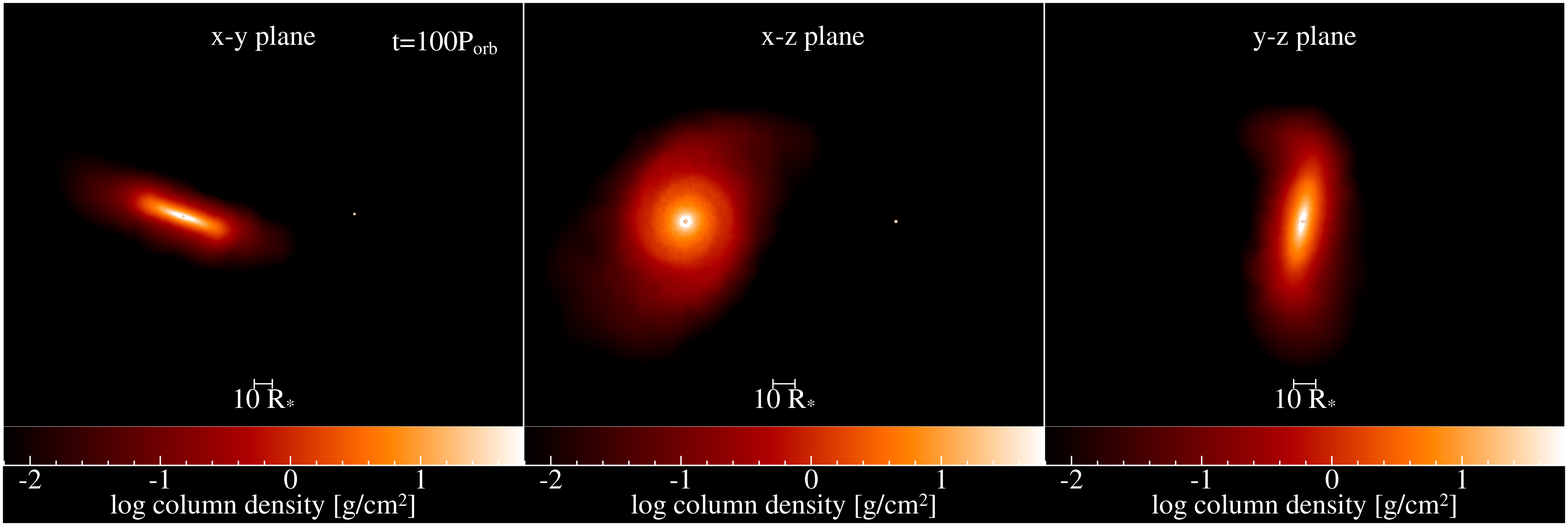}
    \includegraphics[scale = 0.3]{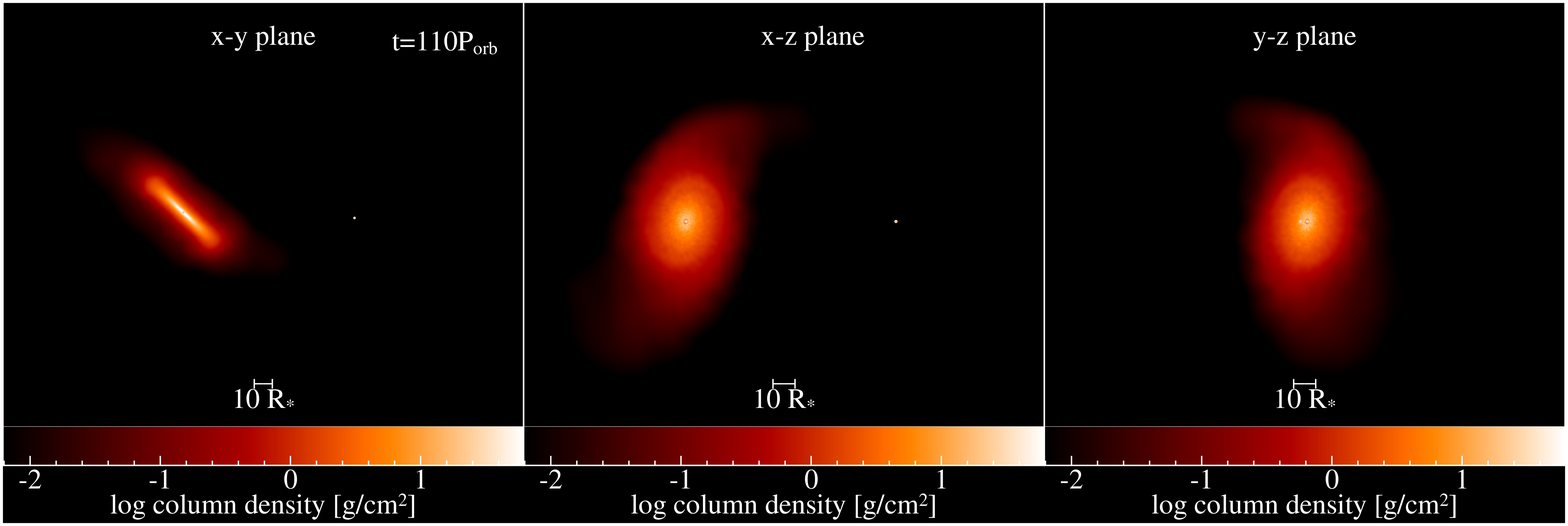}
    \includegraphics[scale = 0.3]{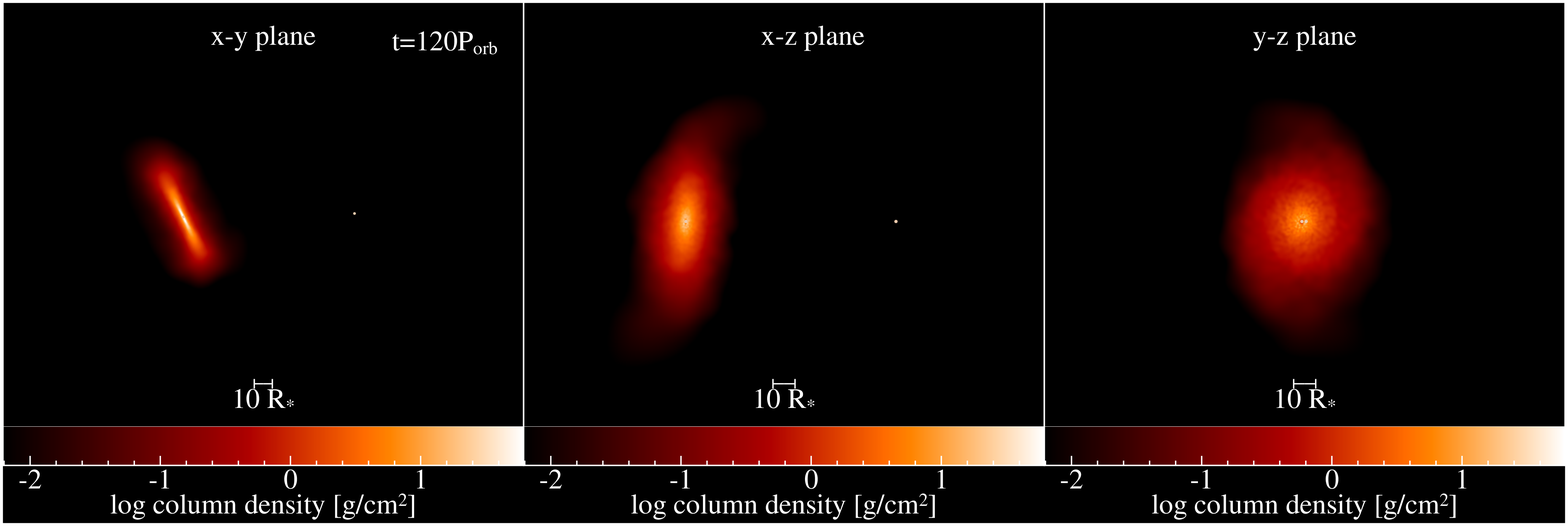}
    \includegraphics[scale = 0.3]{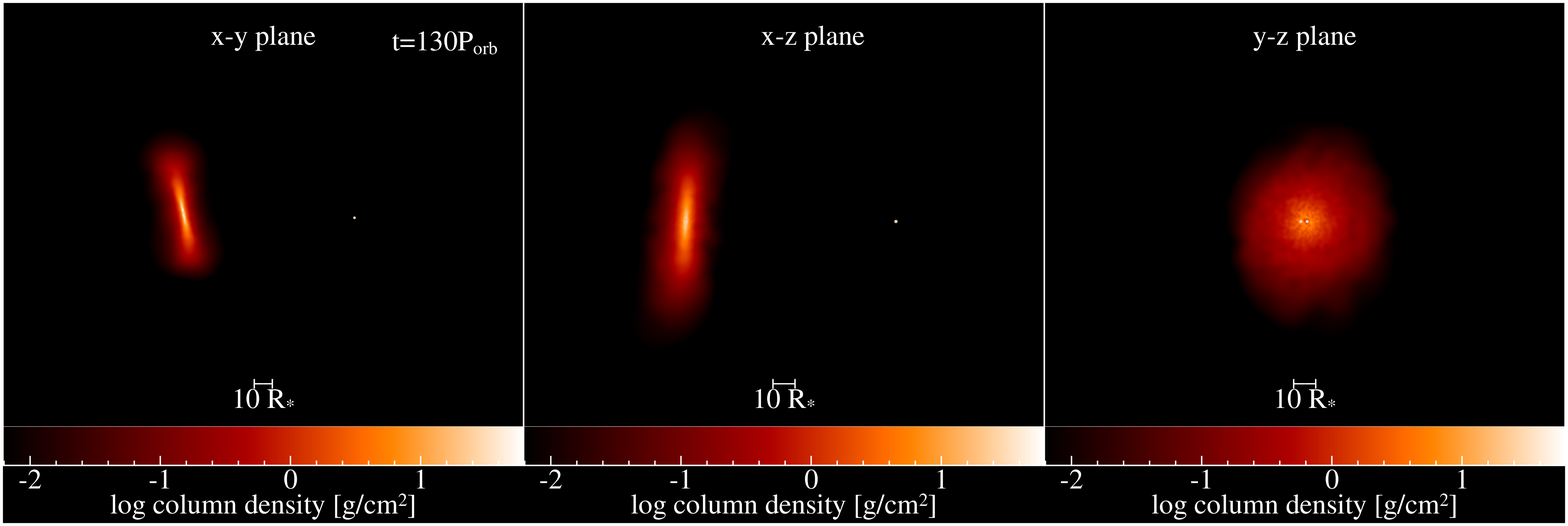}
    \caption{Snapshots of the 300 day, \ang{60} simulation at (top to bottom) 100, 110, 120, and 130 \Porb. Format is the same as Figure \ref{fig:30/40_35Porb}.}
    \label{fig:300/60_100-130Porb}
\end{figure*}

Finally, we show in Figure \ref{fig:300/60_100-130Porb} that, during dissipation, the precession of the disc about the binary's orbital axis can cause substantial changes in the appearance of the disc in the same manner as the 30 day orbital period simulations. The precession periods were found to be approximately 31 and 50 \Porb \text{ }for the \ang{20} and \ang{40} simulations respectively. The precession period for the 300 day, \ang{60} simulation could not be determined due to the disc dissipating before one precession period occurred.

\subsection{Steady-State Disc Structure}
\label{sec:SS_disc_struct}

In all simulations, aside from the 30 day, \ang{40} case where the disc tears, the Be star disc does not undergo considerable changes after 50 to 60 \Porb, up to the start of disc dissipation at 100 \Porb. To analyze the structure during this time, we fit the surface density of the discs using the broken power-law equation
\begin{equation}
    \Sigma (r) = A \frac{(r/R_t)^{-m}}{1+(r/R_t)^{n-m}},
    \label{eq:sigma_fit}
\end{equation}
where $n$ and $m$ are the power-law exponents for the inner and outer part of the disc, and $R_t$ is what we refer to as the transition radius. This formula has been used successfully in previous SPH studies of Be stars \citep{Okazaki2002, panoglou2016discs, Cyr2017}. These previous studies have denoted $R_t$ as the ``truncation radius", however we do not feel this term is appropriate, as there is still significant disc material beyond this point. Our use of the term ``transition radius" better signifies that this point is simply where the material transitions from being dominated by viscous forces, to being more heavily influenced by the binary companion.

We calculate the surface density of the discs using the formula
\begin{equation}
    \Sigma = \int \rho dz.
    \label{eq:sigma_defn}
\end{equation}
To account for the inclination of the disc when calculating the surface density, we rotate the disc about its line of nodes (where it intersects the equatorial plane of the primary star) by its average inclination so that it is approximately centered about the x-y plane, and the integration along the z-axis is minimally skewed by the inclination and warping of the disc.

Since the misaligned binary companion causes asymmetric disc structures, we choose not to fit the azimuthally averaged surface density as past studies have done. Instead, we fit the surface density at separate azimuthal angles around the disc, out to the point where the surface density reaches $10^{-10}\,\rm g\,cm^{-2}$. This fitting is done at timesteps of 95, 96, 97, 98 and 99 \Porb, and then the fitting parameters ($R_t$, $n$, and $m$) are averaged over the five fits for each azimuthal angle to reduce any noise that may result from fitting a single timestep.

\begin{figure}
    \centering
    \includegraphics[scale = 0.33]{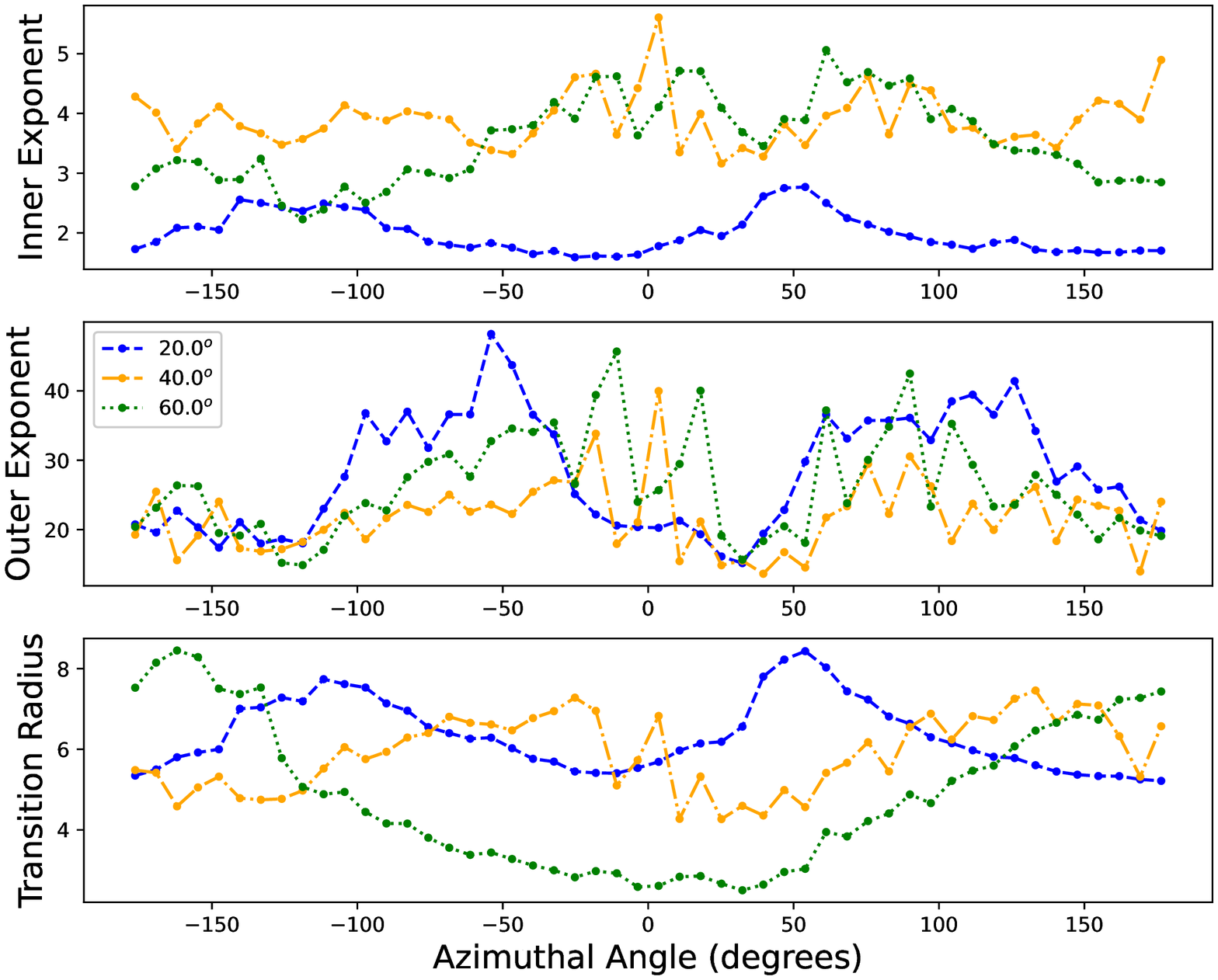}
    \caption{Top to bottom, the average fits of the inner exponent, outer exponent, and transition radius as defined in Equation \ref{eq:sigma_fit} to the surface density at 5 timesteps from 95 to 99 \Porb, for the 30 day simulations. The x-axis denotes the azimuthal angle where the surface density was fit. As indicated by the legend, the \ang{20}, \ang{40}, and \ang{60} simulations are shown as the dashed blue, dashed-dotted orange, and dotted green line respectively.}
    \label{fig:30_sigma_fits}
\end{figure}

\begin{figure}
    \centering
    \includegraphics[scale = 0.33]{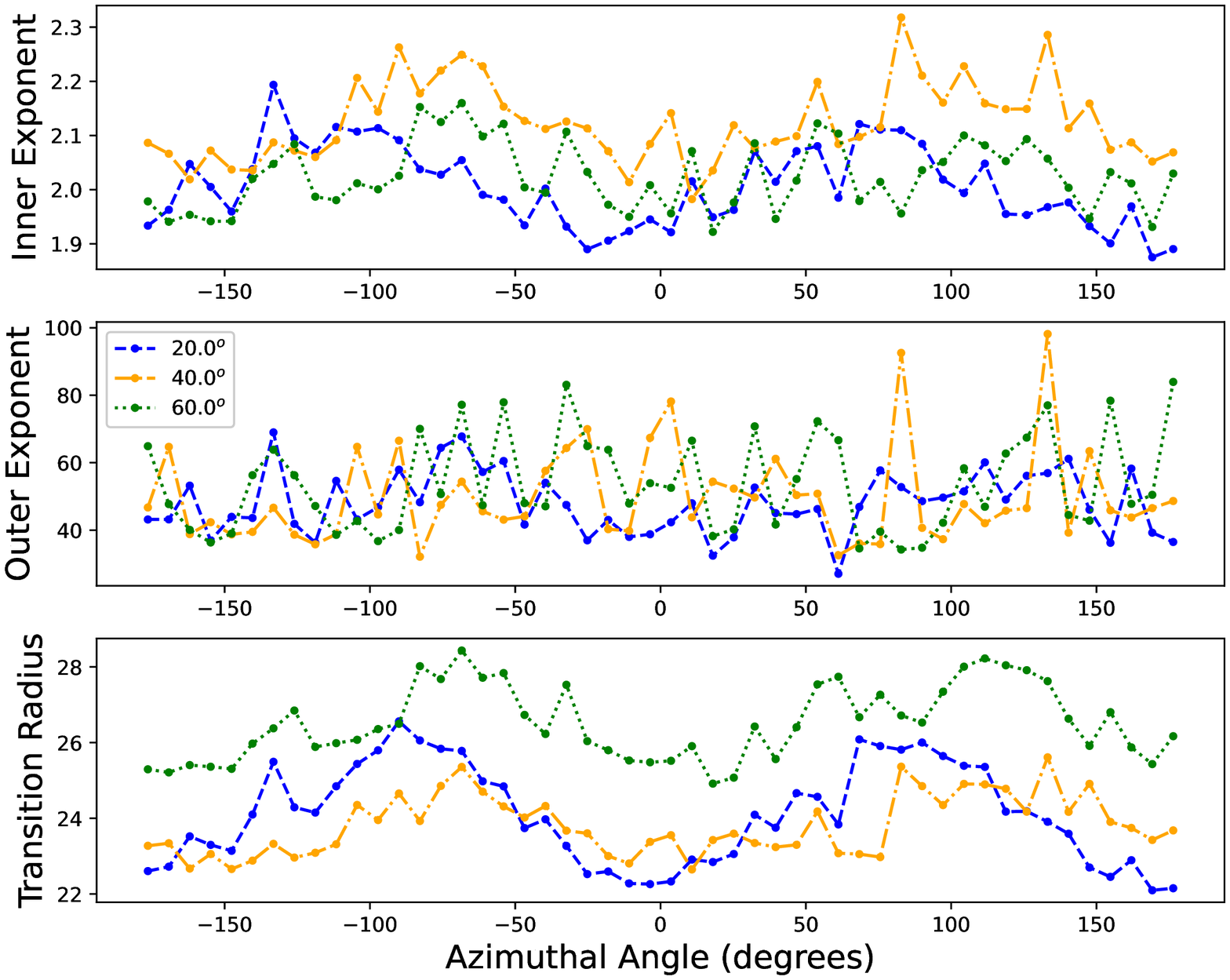}
    \caption{Same format as Figure \ref{fig:30_sigma_fits}, but for the 300 day simulations.}
    \label{fig:300_sigma_fits}
\end{figure}

Figures \ref{fig:30_sigma_fits} and \ref{fig:300_sigma_fits} show the results of this fitting process for the 30 day and 300 day simulations, respectively. Note that the results for the 30 day, \ang{40} simulation contain extra uncertainty due to the disc being at the end of a recombining period after tearing into separate ring and disc components. Other noteworthy results here are in the 30 day, \ang{60} simulation, where due to the highly elliptical and asymmetric shape of the disc, the transition radius varies drastically from an azimuthal angle of \ang{0} to \ang{180}. While the other discs also show ellipticity in their truncation radii, the 30 day, \ang{60} model is the only to show this drastic disc asymmetry, with all other discs having two maxima and two minima in their truncation radii around the disc.

\subsection{Radial Variations in Disc Eccentricity}
\label{sec:disc_ellipse}

As mentioned previously, the 30 day, \ang{60} simulation is the only simulation to show a considerable disc eccentricity when averaging over the whole disc. However as noted in Section \ref{sec:SS_disc_struct}, we also notice that the discs of other simulations do not appear to have circular symmetry when looking at their transition radii, despite their average eccentricity being nearly zero. To explain this difference, at a certain timestep in our simulations, for each particle we calculate its instantaneous orbital parameters using its position and velocity. We then sort the particles into 100 particle bins according to increasing distance from the primary star and calculate their average orbital eccentricity in each bin.

\begin{figure*}
    \centering
    \begin{subfigure}[b]{0.3\textwidth}
         \centering
         \includegraphics[width=\textwidth]{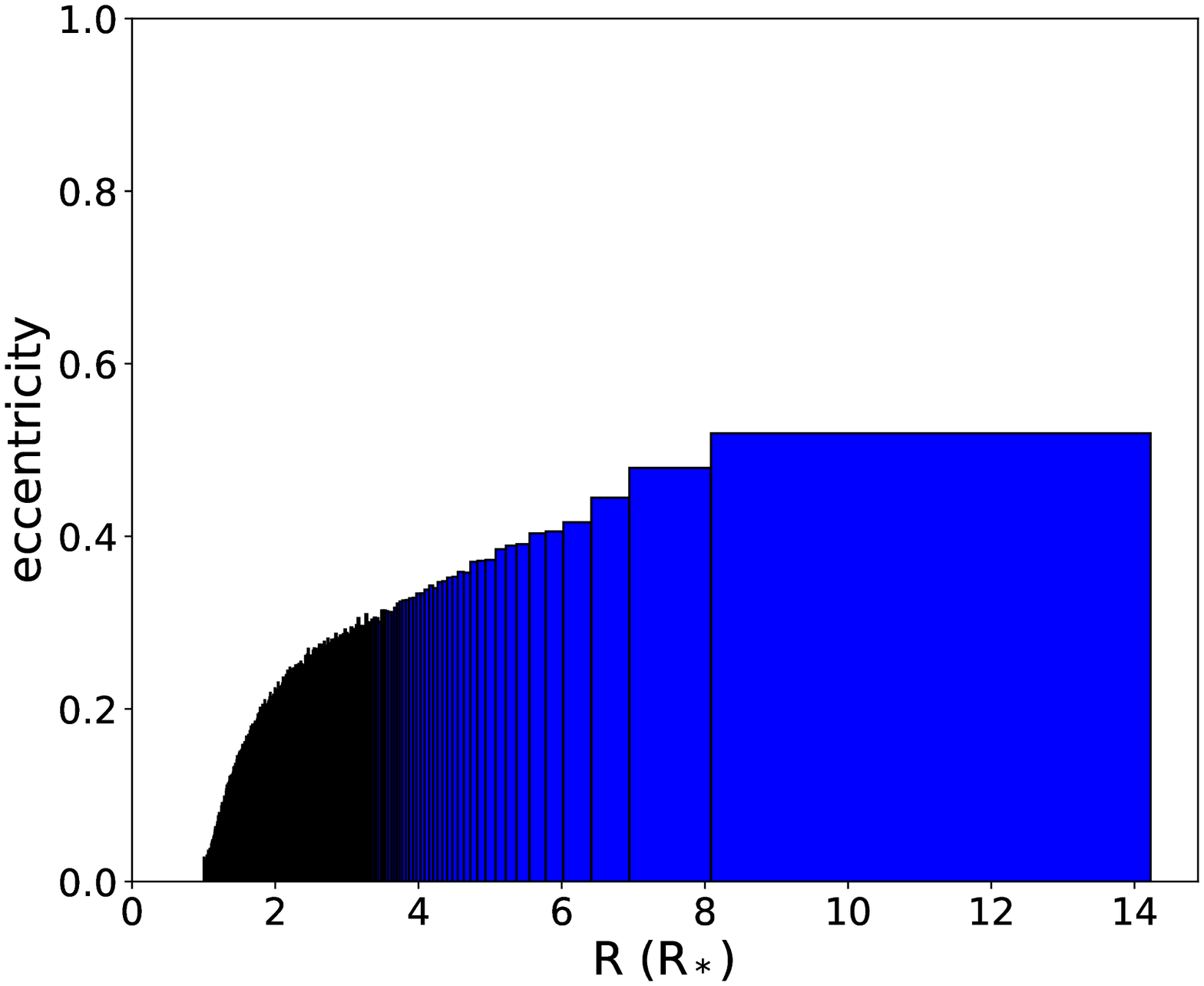}
         \caption{30 day, \ang{60}}
         \label{fig:30/60_bar_ell}
     \end{subfigure}
     \begin{subfigure}[b]{0.3\textwidth}
         \centering
         \includegraphics[width=\textwidth]{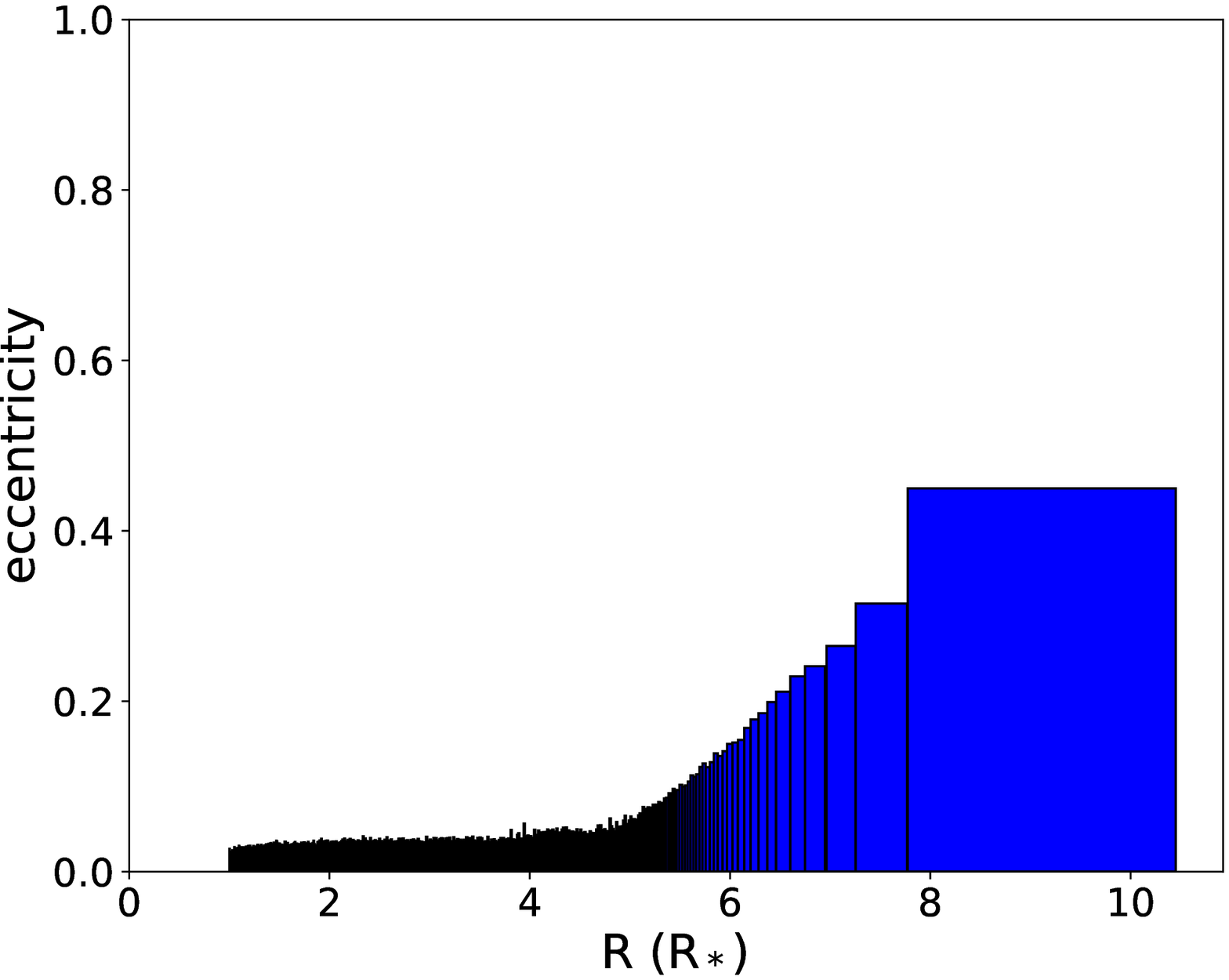}
         \caption{30 day, \ang{20}}
         \label{fig:30/20_bar_ell}
     \end{subfigure}
     \begin{subfigure}[b]{0.3\textwidth}
         \centering
         \includegraphics[width=\textwidth]{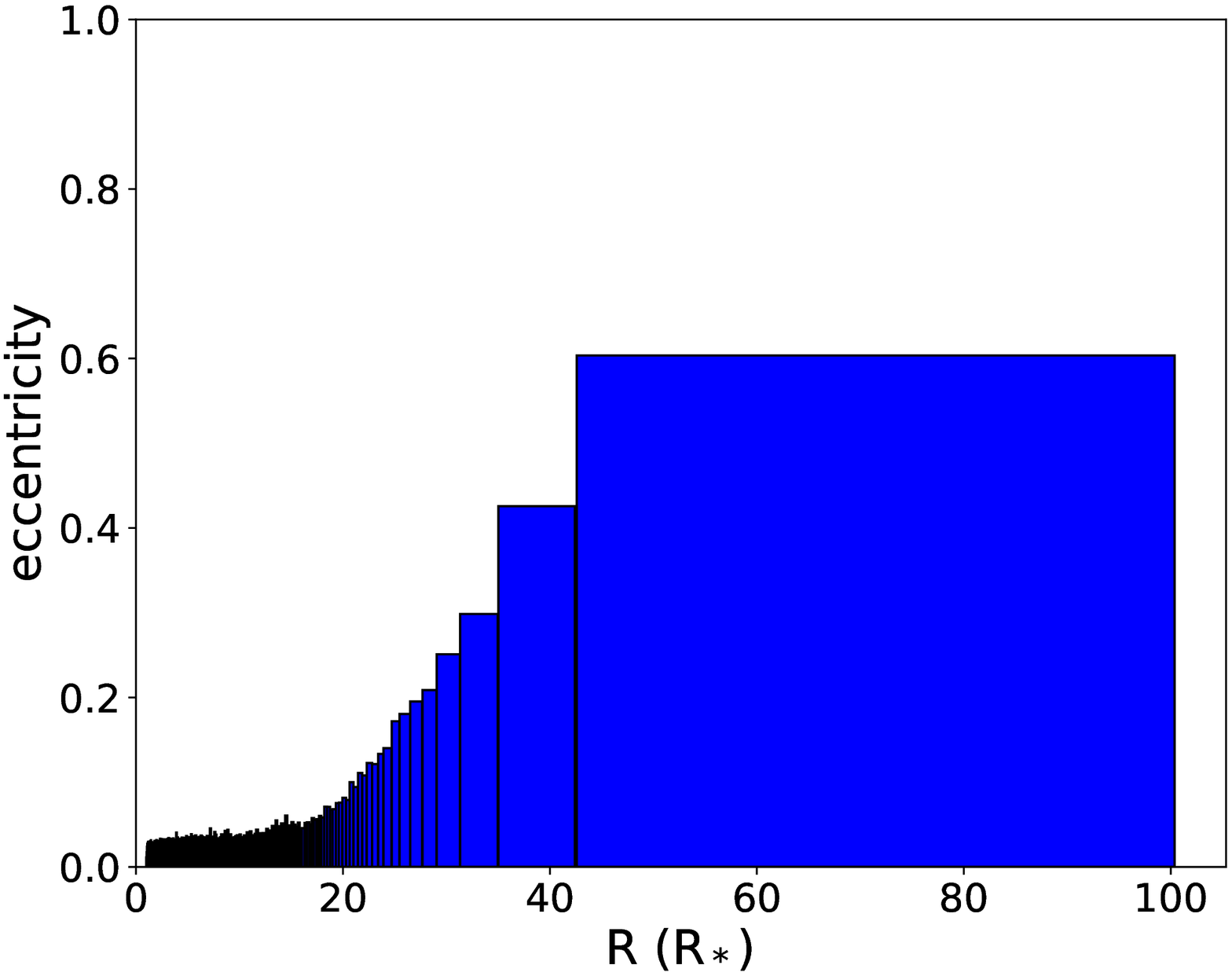}
         \caption{300 day \ang{20}}
         \label{fig:300/20_bar_ell}
     \end{subfigure}
    \caption{Average eccentricity of 100 particle bins at 95 \Porb \text{ } of (a) our 30 day, \ang{60} simulation, (b) our 30 day, \ang{20} simulation, and (c) our 300 day, \ang{20} simulation. The particles are binned according to their distance from the primary star. The x-axis limits of each bin is the radial position of the closest and farthest particle in each bin. Note that there may be less than 100 particles included in the last bin.}
    \label{fig:bar_ellipses}
\end{figure*}

Figure \ref{fig:bar_ellipses} shows this analysis plotted for our 30 day, \ang{60}, 30 day, \ang{20}, and 300 day, \ang{20} simulations. As expected, due to the high average eccentricity of the 30 day, \ang{60} simulation, the eccentricity of the particles rises quickly and levels off with distance from the primary star. The other two simulations however, are surprising in that the disc particles have very circular orbits in the inner disc, but the outer disc becomes increasingly eccentric, up to a maximum of 0.4 to 0.6. As shown in both Figures \ref{fig:30/20_bar_ell} and \ref{fig:300/20_bar_ell}, the particles eccentricity starts to dramatically increase at or just before the transition radius of the disc. This same pattern also occurs for the other simulations, both long and short period, not pictured in Figure \ref{fig:bar_ellipses}. The final bin in all three graphs are large due to some particles that may become disassociated with the rest of the disc and are being drawn towards the binary companion.

\begin{figure*}
    \centering
    \begin{subfigure}[b]{0.4\textwidth}
         \centering
         \includegraphics[width=\textwidth]{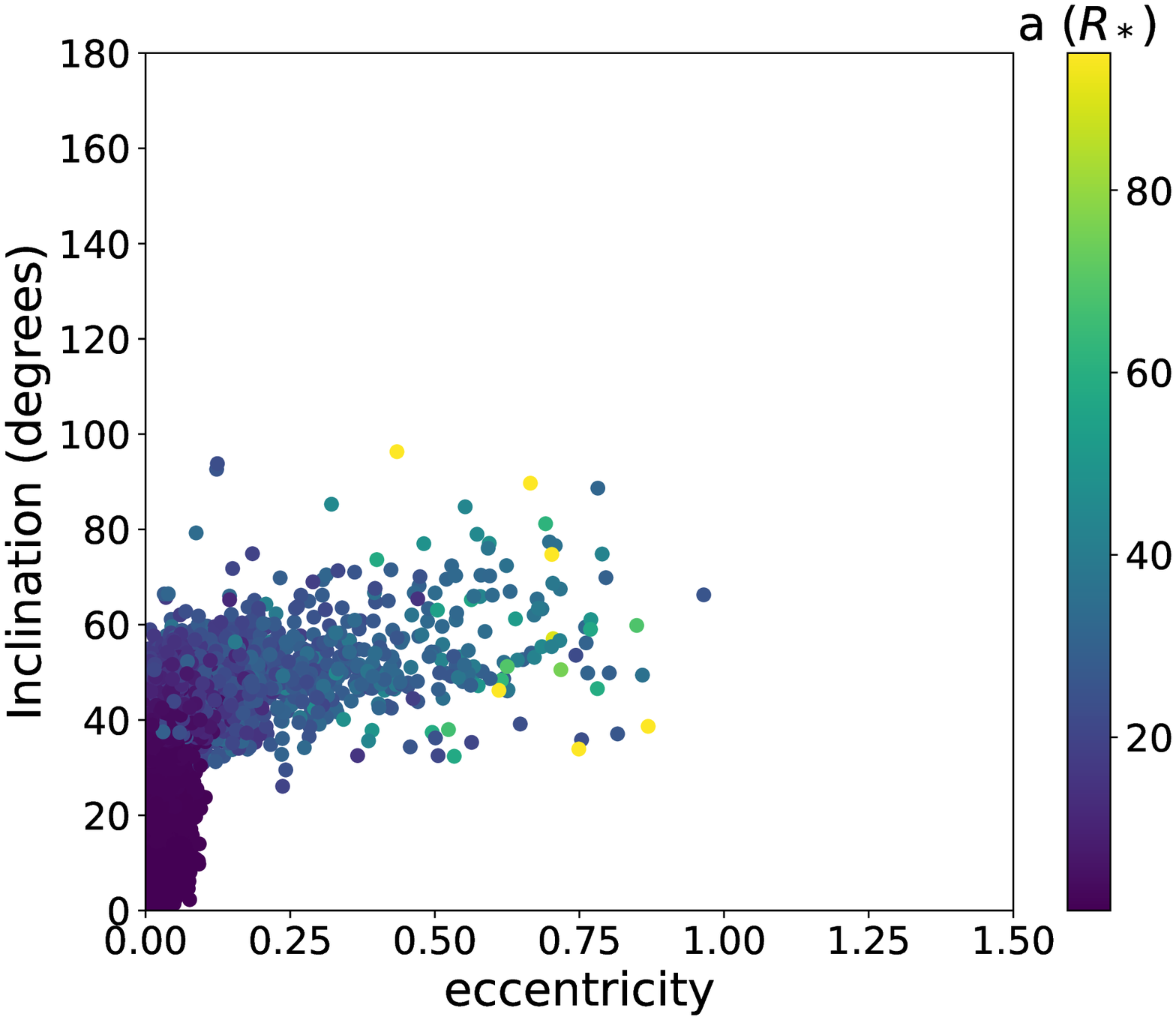}
         \caption{300 day, \ang{40}, 100 \Porb}
         \label{fig:300/40/100_part_scatter}
     \end{subfigure}
     \begin{subfigure}[b]{0.4\textwidth}
         \centering
         \includegraphics[width=\textwidth]{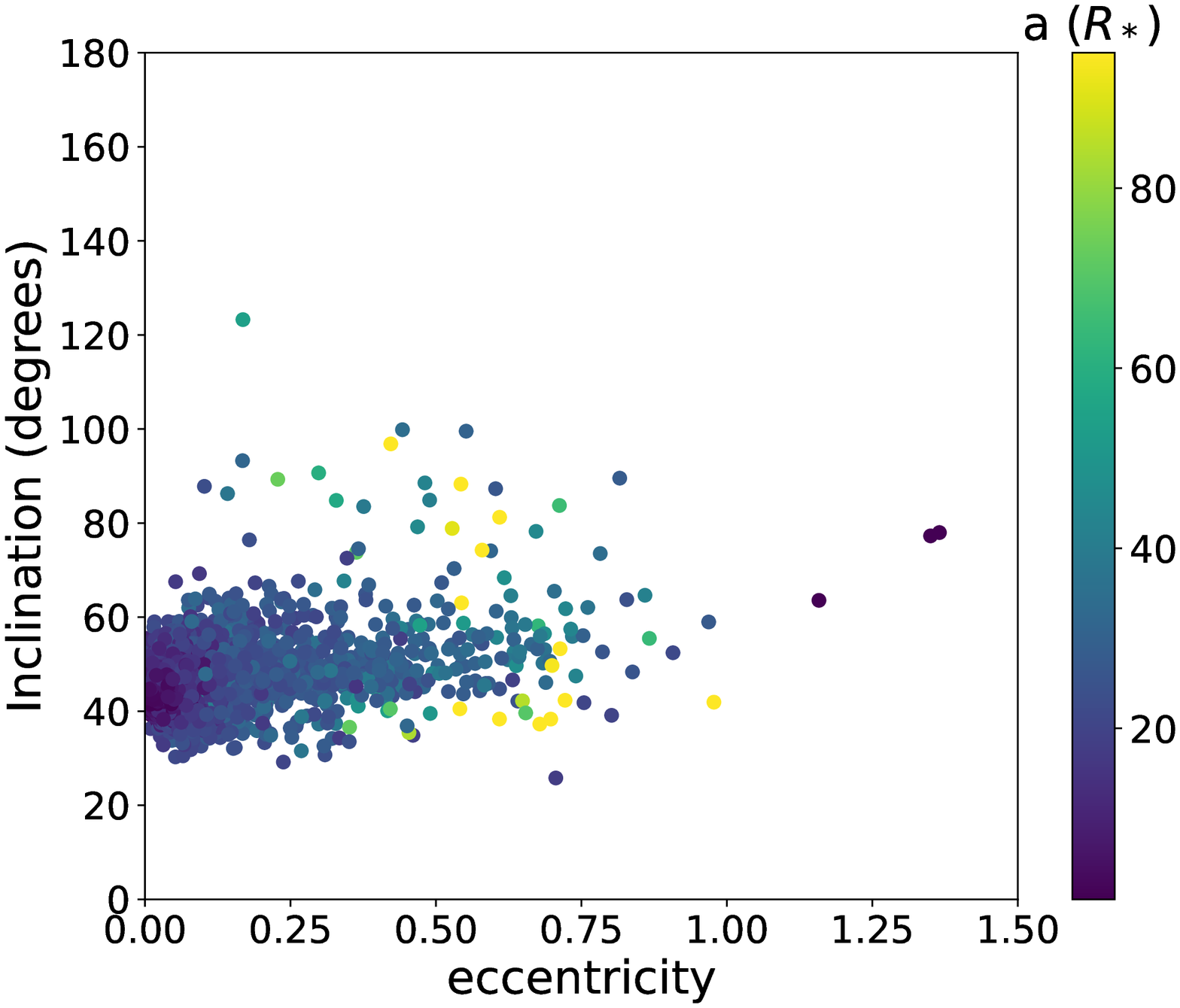}
         \caption{300 day, \ang{40}, 101 \Porb}
         \label{fig:300/40/101_part_scatter}
     \end{subfigure}
     \\
     \begin{subfigure}[b]{0.4\textwidth}
         \centering
         \includegraphics[width=\textwidth]{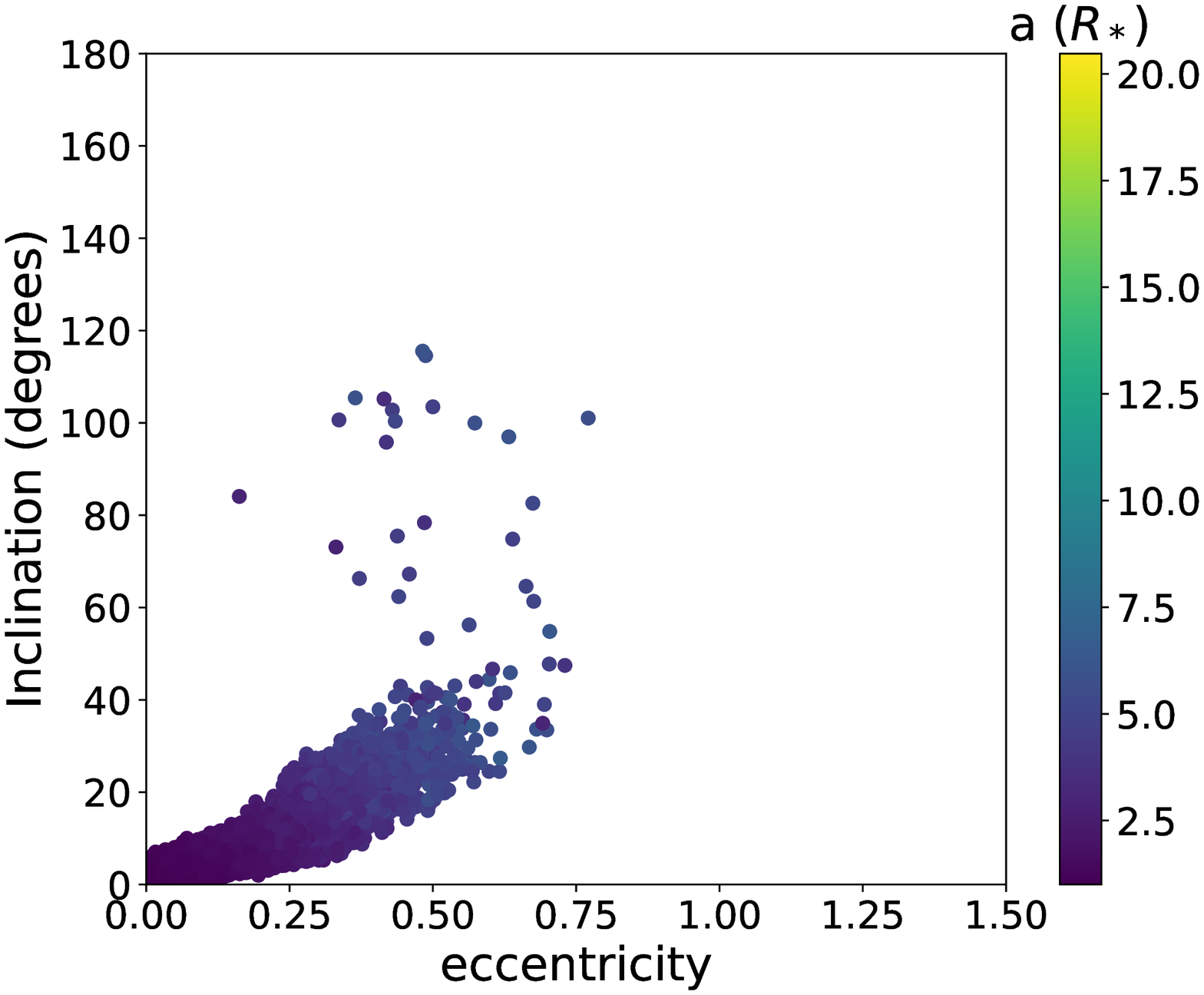}
         \caption{30 day, \ang{60}, 100 \Porb}
         \label{fig:30/60/100_part_scatter}
     \end{subfigure}
     \begin{subfigure}[b]{0.4\textwidth}
         \centering
         \includegraphics[width=\textwidth]{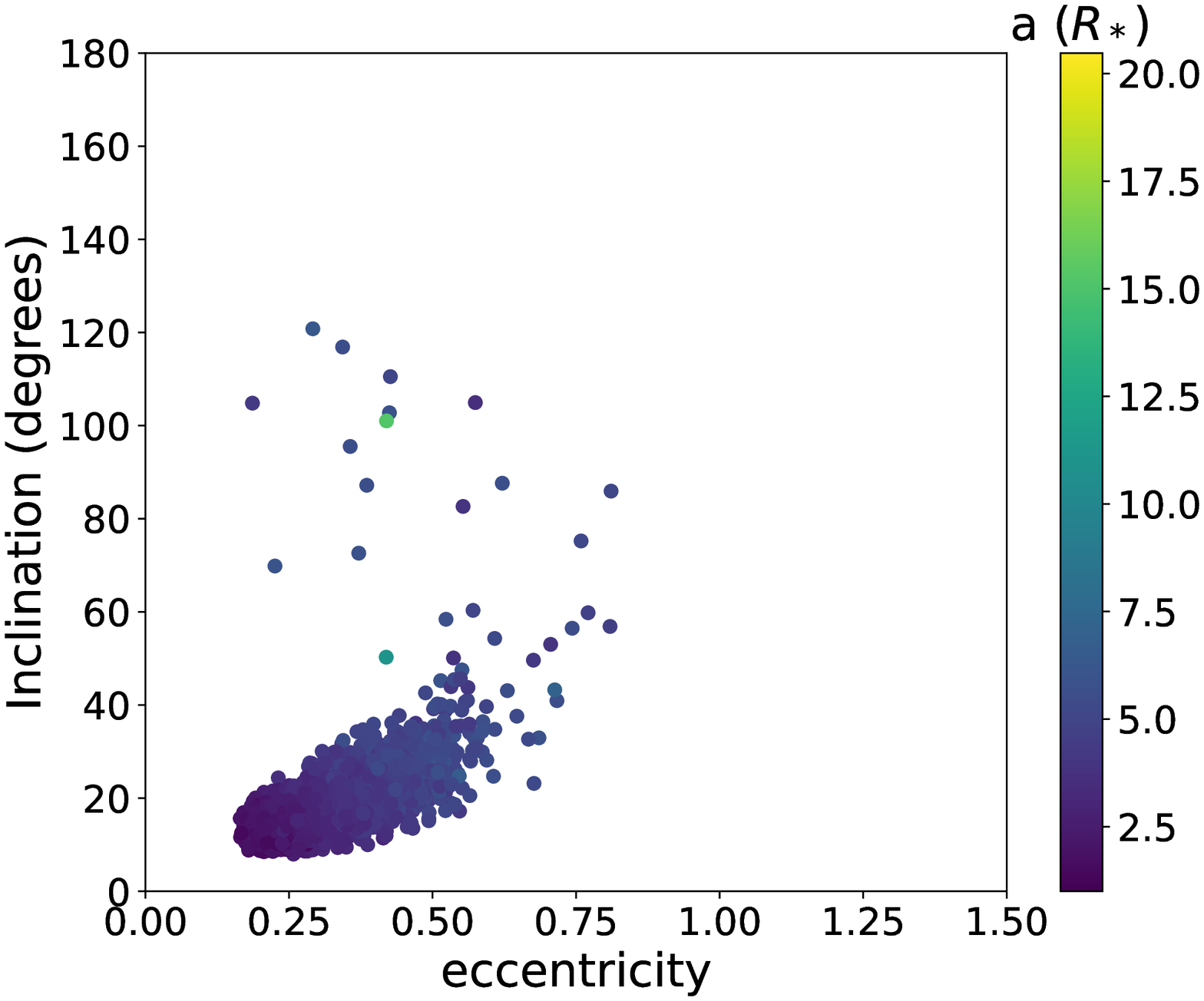}
         \caption{30 day, \ang{60}, 101 \Porb}
         \label{fig:30/60/101_part_scatter}
     \end{subfigure}
    \caption{Plots of individual particle eccentricity and inclination at 100 \Porb \text{ }and 101 \Porb \text{ } for the 300 day, \ang{40} simulation (top row) and the 30 day, \ang{60} simulation (bottom row). The particles are coloured according to their orbital semi-major axis, denoted by the colour bar in each subfigure.}
    \label{fig:part_scatter_incl_ecc}
\end{figure*}

The particle eccentricity also has a relationship with its inclination. Figure \ref{fig:part_scatter_incl_ecc} shows scatter plots of individual particle eccentricity versus inclination for our 300 day, \ang{40} simulation, and our 30 day, \ang{60} simulation, at at 100 (at the end of the active phase) and 101 (just after mass injection is turned off) \Porb. In the 300 day, \ang{40} simulation, we see that at 100 \Porb, most particles are at a low eccentricity, but have a large spread in inclination. The more eccentric particles, however, have large non-zero inclinations. These highly eccentric, highly inclined particles are also those that are furthest from the primary star, as indicated in the Figure by their semi-major axis. After dissipation begins, we see in Figure \ref{fig:300/40/101_part_scatter} that the particles closest to the primary star, with low inclination and low eccentricity, quickly reaccrete onto the primary star, and only the highly inclined and eccentric particles remain. This indicates that all phenomena described earlier during dissipation (KL oscillations, disc precession, etc.) happens in the outermost part of the disc.

The overall characteristics of Figures \ref{fig:300/40/100_part_scatter} and \ref{fig:300/40/101_part_scatter} are seen in all other simulations as well, except for the 30 day, \ang{60} simulation, which is shown in Figures \ref{fig:30/60/100_part_scatter} and \ref{fig:30/60/101_part_scatter}. Here, we see that the inclination and eccentricity of the particles increase smoothly together with increasing distance from the primary star. This points to the large average disc eccentricity seen in this simulation found in Section \ref{sec:30/60}. Once again after disc dissipation begins, the inner particles with the lowest eccentricity and inclination immediately reaccrete, leaving only highly eccentric and inclined particles.

\section{Discussion}
\label{sec:Discussion}

\subsection{Disc Evolution}
\label{sec:Discuss_evol}

The simulations presented here are a critical step towards predicting how Be star observables may change due to the effect of a misaligned binary companion. The simulations involve an equal-mass binary system with constant viscosity parameter $\alpha$, while varying the binary orbital period and binary misalignment angle. The Be star disc growth and subsequent dissipation were simulated by having the mass-loss turned on at a constant rate of $10^{-8} \, \rm M_\odot yr^{-1}$ in the simulation for 100 \Porb, and then turning the mass-loss off at 100 \Porb \text{ }to allow the disc to dissipate and reaccrete onto the primary star. Note that while $\alpha$ has been shown to merely act as a time-scaling parameter in single-star Be systems \citep{Haubois2012}, a different $\alpha$ in a binary system may cause much different disc evolution that what we have presented here.

The basic behaviour of the discs in all simulations is as follows. During the disc growth phase the outer disc is tilted away from the equatorial plane due to the misaligned companion, before settling into a steady configuration until the mass-injection is turned off at 100 \Porb. Then, during disc dissipation, the disc precesses about the binary's orbital axis. Disc precession is not surprising during the dissipation phase, since at this stage the disc can be considered an accretion disc, with the accretion region growing larger with time \citep{Haubois2012}. Disc precession has been found in accretion discs by \cite{Larwood1996} and \cite{Dogan2015}, for example. Equation 21 of \cite{Larwood1996} gives a formula for the precession period of a rigid disc, which is dependent on the misalignment angle of the binary, the surface density of the disc, and other constant parameters. Using the azimuthally averaged surface density for our discs at 100 \Porb \text{ }(the start of disc dissipation) we find this formula predicts precession periods that are within a factor of two to the periods we find from our simulations. Given the equation is meant for rigid discs, and we used the azimuthal average of the surface density, this is an encouragingly close result to the precession periods found from our simulations.

There is one other interesting finding that appears in all simulations, except for the 30 day, \ang{20} simulation which is least affected by the misaligned binary, where during the tilting of the disc in the disc growth phase, the mass of the disc oscillates in the same manner as the inclination. This same occurrence was predicted in the analytical models of \cite{Martin2011}. Mini-dissipation phases like this while the mass-injection is still on could have large implications to the conservation of angular momentum of the system and rotation rate of the primary star, as well as effect the system's observables.

The most unusual results came from the simulations with a 30 day orbital period, and \ang{40} or \ang{60} misalignment angles. During the disc growth phase, the 30 day, \ang{40} simulation underwent periodic disc tearing episodes where an outer ring separated from the inner disc and precessed independently before recombining with the rest of the disc. Similar behaviour has been shown to occur in accretion discs by \cite{Dogan2015}, however the interesting difference here is that these tearing episodes occur during the disc growth phase in our simulations, before the mass-loss rate is turned off and the disc is allowed to dissipate. So it appears that disc tearing is not just confined to an accretion disc scenario. This is the only simulation in our investigation where we see this phenomenon take place. The other simulations likely do not satisfy the condition for disc tearing where precession torque is required to be greater than the disc's viscous communication \citep{Dogan2015}.

The 30 day, \ang{60} simulation presents entirely different behaviour from any other of our simulations. During the disc growth phase, the disc becomes moderately eccentric, eventually settling to an average eccentricity of about 0.2. The disc is also highly asymmetric, with the shortest side of the disc only extending radially outward about a quarter of the distance of the longest side. Once disc dissipation begins, an eccentric gap opens between the primary star and disc, a feature not present in any other of our simulations. Be star discs are thought to gradually lose their disc from the inside out, with the inner density dropping quickly and being filled in from material further out in the disc, though not to the extent of forming a ring-like structure as seen here \citep{rivinius2013classical}. 

While this gap is present, we see the disc undergo KL oscillations until dissipation is complete. The oscillations seen here are very similar to those shown by \cite{Martin2014} in that they are damped, while traditional KL oscillations involving a test particle would have a constant amplitude. The disc flipping to a retrograde orbit with respect to the binary's orbit is also a feature known to be possible in KL oscillations \citep{Naoz2013}. The analytical KL timescale calculated in Section \ref{sec:30/60} is in good agreement with our initial observed KL timescale. This of course is only valid for the first oscillation, since the surface density of the disc evolves as dissipation progresses. As shown in Section \ref{sec:30/40}, we also detect KL oscillations at the very end of our 30 day, \ang{40} simulation. It seems appropriate that these are the only simulations in this work where we detect these oscillations, as the minimum inclination needed between the binary and inner orbiting object for these oscillations to occur is \ang{39.2} for a test particle, and has been shown to be able to be lower in discs \citep{Lubow2017}. \cite{Lubow2017} find that the possibility for KL oscillations in discs is more dependent on the disc aspect ratio (scale height over radius) than the inclination difference with the binary, so the discs in our other simulations most likely exist outside of this required instability range of aspect ratios for these oscillations to occur. Thus it is likely that KL oscillations will occur in a variety of Be star systems as the disc dissipates, and will occur at different times depending on the parameters of the system. Due to the nature of our simulations, where we grow the disc from nothing and allow it to evolve over time, the aspect ratio is not a set parameter and is very difficult to estimate given the tilted and warped disc structures in our simulations. So because of this uncertainty, we omit this specific analysis from this paper. 

\subsection{Steady-State Disc Properties}
\label{sec:discuss_SS}

As mentioned previously, for some time after the disc has grown, but before disc dissipation occurs, the disc in every simulation (except for the 30 day, \ang{40} case) achieves a near steady configuration. As shown in Section \ref{sec:SS_disc_struct}, in this steady configuration, the disc density is severely modified by the companion star all around the disc at the transition radius. Fitting the surface density with a broken power-law (Eq. \ref{eq:sigma_fit}), we find that this transition point always occurs at a radius within the Roche Lobe, and that the outer density exponent is tens of times larger than the inner exponent. If the mass ratio was lower, this effect would not be as severe, and the outer exponent would be much closer to the inner exponent such as those found by \cite{Cyr2017}. As discussed by \cite{panoglou2016discs} and \cite{Cyr2017}, raising or lowering $\alpha$ or the mass-loss rate would respectively result in larger or smaller discs.

The eccentricity of the disc was also found to vary with radius during the steady-state phase. From the transition radii in Figures \ref{fig:30_sigma_fits} and \ref{fig:300_sigma_fits}, we can see that for most discs, there are two maxima and two minima on opposite sides of the disc, indicating an elliptical disc shape. We further showed this in Section \ref{sec:disc_ellipse} by averaging eccentricity of the particles in 100 particle bins, and find for all simulations, except the very eccentric 30 day, \ang{60} case, that the disc inward of the transition radius is in nearly circular orbit, while near and farther out from the transition radius the particle orbits become quite eccentric. This may be due to the severe drop off of density near the transition radius, which would reduce viscous effects in the disc and allow the particles to be more heavily influenced by the binary companion. Future work running simulations involving more particles, and thus a higher resolution in the outer disc will be able to further show these effects in more detail. We also showed, in Figure \ref{fig:part_scatter_incl_ecc}, that in general, particles with high eccentricity have a high inclination, and are farther away from the primary star. This pattern does not reciprocate, however, as highly inclined particles can still have low eccentricities, and have relatively small semi-major axes.

\subsection{Effect on Observables}
\label{sec:discuss_observe}

Although simulated observations of these models will not be presented here, and instead will be in a follow up paper to this study to allow the fully detailed results to be presented, the findings in these models give us great reason to believe a misaligned binary companion will have a large effect on all Be star observables.

Depending on the disc's inclination to the observer, photometric measurements of a Be star may increase or decrease during the growth and dissipation phases \citep{Haubois2012}. However a change in disc inclination will also affect photometric magnitudes, dimming as the disc moves to be edge-on, and brightening as the disc transitions to a more pole-on orientation. Depending on the initial view of the observer, the changes in disc inclination seen in our models will definitely be able to be seen as either a brightening or dimming throughout the simulation. The photometric magnitudes may even oscillate back and forth, for example as the disc precesses during its dissipation phase.

Following the same reasoning, the Balmer emission lines will also be seen to transition from shell phase at edge-on orientations, to Be phase at pole-on orientations \cite[for an example of such behaviour, see the Be star Pleione;][]{Hirata2007}. Again we predict these lines will oscillate back and forth as the disc oscillates in inclination.

Perhaps the most striking observations of a changing disc orientation will be seen in the linear polarization measurements produced from the disc. Specifically, if the disc is precessing, the polarization position angle should also oscillate with the disc, again as seen in Pleione \citep{Hirata2007}. However, since the precession in our models occurs in the dissipation phase and the polarization is largely produced in the inner disc \cite[see figure 1 of][]{Carciofi2011}, it will be interesting to see if the polarization signature is strong enough from the remaining disc to detect significant changes as the disc orientation changes.

Though KL oscillations have yet to be detected in Be star discs, the behaviour seen in our 30 day, \ang{60} simulation will allow us to quantify an observational signature of these oscillations. The observables mentioned above should all be affected by KL oscillations, however it may produce a different signature than normal disc precession seen in our other simulations. This will surely have application to KL oscillations seen in some types of accretion discs as well. Of course, the changes in observables resulting from these simulations cover only a small subset of system parameters. A change in $\alpha$, the mass-loss rate, or the mass ratio of the system should also be explored in future work to add to our knowledge of the possible effects of a misaligned binary companion.

\section{Conclusion}
\label{sec:conclusion}

The simulations presented in this work show new and interesting behaviour during the growth and dissipation phases of Be star discs in misaligned binary systems. Through 3D SPH simulations, we simulated equal-mass binary systems with a circular binary orbit of either 30 or 300 days, inclined to the equatorial plane of the primary star by either \ang{20}, \ang{40}, or \ang{60}.

Our work shows that, in general, a misaligned binary companion can cause significant tilting of the disc while it is growing, and that this tilting can cause the disc to partially reaccrete onto the primary star while the mass-injection into the disc is still active, which is seen in a majority of our simulations. During disc dissipation, the disc transitions to a pure accretion disc, and in almost all cases, regardless of orbital period, precesses about the binary companion's orbital axis with precession periods between 20 and 50 \Porb. We have also shown that a misaligned companion can result in a steady asymmetric disc structure that is greatly truncated due to the binary companion, and that the outer parts of the disc can have far more eccentric and inclined orbits than the inner portions.

The results of these simulations also show that phenomena such as disc tearing and KL oscillations are able to occur in Be star discs. Disc tearing occurs in our 30 day, \ang{40} simulation while the mass-injection is still on in the simulation. The disc periodically tears into two separate discs and merges back into one disc approximately every 30 \Porb. This shows the tearing phenomenon is not just confined to accretion discs. We detect damped KL oscillations in our 30 day, \ang{60}, and 30 day, \ang{40} simulations after the mass-injection into the disc is turned off. The eccentricity and inclination of the disc in the \ang{60} case oscillate on approximately 5 \Porb \text{ } periods, reaching a maximum eccentricity of about 0.6, and a minimum inclination of about \ang{20} with respect to the binary's orbit. The KL oscillations seen in the \ang{40} case begin in the final 10 orbital periods of the simulation, thus no period or maximum and minimum numbers could be determined. Neither disc tearing or KL oscillations are detected in any other of our simulations, however the difference in timing of the onset of KL oscillations in our two detections shows that these phenomena likely can occur in a variety of different binary Be star configurations.
 
It will be of great interest in our next work to use the results of these simulations in a radiative transfer code, to predict observables from these disc structures, which will allow us to quantify the observable effect a misaligned binary companion would have on a Be star and its disc. These results will help determine the prevalence of Be stars in misaligned binary systems, as well as give clues to the formation of Be disc systems.

\section*{Acknowledgements}

We thank the anonymous referee for their very enthusiastic and helpful comments which improved the paper. C.E.J. acknowledges support through the National Science and Engineering Research Council of Canada. A. C. C. acknowledges support from CNPq (grant 311446/2019-1) and FAPESP (grants 2018/04055-8 and 2019/13354-1). This work has made use of the computing facilities of the Laboratory of Astroinformatics (IAG/USP, NAT/Unicsul), whose purchase was made possible by the Brazillian agency FAPESP (grant 2009/54006-4) and the INCT-A. We acknowledge the use of {\sc splash} \citep{price_2007} for rendering and visualization of our figures. This work was made possible through the use of the Shared Hierarchical Academic Research Computing
Network (SHARCNET).

\section*{Data Availability}

No new data were generated or analysed in support of this research.

\bibliographystyle{mnras}
\bibliography{aastexBeStarbib}

\bsp	
\label{lastpage}
\end{document}